\documentclass[journal]{IEEEtran}

\usepackage{cite}
\usepackage{amsmath,amssymb,amsfonts}
\usepackage{algorithmic}
\usepackage{graphicx}
\usepackage{textcomp}
\usepackage{caption}
\usepackage{subcaption}
\usepackage{float}
\usepackage{array}
\usepackage{xfrac}
\usepackage{xcolor}
\usepackage{relsize}

\usepackage{tikz}

\newcommand\submittedtext{%
  \footnotesize © 2026 IEEE. All rights reserved, including rights for text and data mining and training of artificial intelligence and similar technologies. Personal use is permitted,
but republication/redistribution requires IEEE permission. See https://www.ieee.org/publications/rights/index.html for more information}

\newcommand\submittednotice{%
\begin{tikzpicture}[remember picture,overlay]
\node[anchor=south,yshift=10pt] at (current page.south) {\fbox{\parbox{\dimexpr0.65\textwidth-\fboxsep-\fboxrule\relax}{\submittedtext}}};
\end{tikzpicture}%
}

\begin{document}

%
\title{Theoretical Studies of Sub-THz Active Split-Ring Resonators for Near-Field Imaging}
%
%
%

\author{Ali Ameri, Jun-Chau Chien, and Ali M. Niknejad}

\maketitle

\submittednotice

\begin{abstract}
This paper develops a theoretical framework for the design of Active Split-Ring Resonators (ASRRs). An ASRR is a Split-Ring Resonator (SRR) equipped with a tunable negative resistor, enabling both switchability and quality factor boosting and tuning. These properties make ASRRs well-suited for integration into dense arrays on silicon chips, where pixelated near-fields are generated and leveraged for high-resolution 2D imaging of samples. Such imagers pave the way for real-time, non-invasive, and low-cost imaging of human body tissue. The paper investigates ASRR coupling to host transmission lines, nonlinear effects, signal flow, and the influence of various noise sources on detection performance. Verified through simulations, these studies provide design guidelines for optimizing the Signal-to-Noise Ratio (SNR) and power consumption of a single pixel, while adhering to the constraints of a scalable array.
\end{abstract}


%
\IEEEpeerreviewmaketitle

\section{Introduction}
%
%
%
%
\IEEEPARstart{N}{ear}-fields enable high-resolution sub-wavelength imaging systems by overcoming the fundamental resolution limitations of far-field imaging, which is constrained to approximately half the wavelength of the excitation signal \cite{huang_diffraction_cell2010}. Unlike far-fields, near-fields can be conveniently generated on a chip surface using compact, pixelated designs. The most common approaches include coplanar electrode arrays at RF and microwave frequencies \cite{hu_imgr_tbiocas2022,cheng_imgr_mwtl2023} and resonant structures at mm-wave frequencies and above \cite{mitsunaka_jssc2016,hillger_jssc2018,tanaka_tcas2022}. 

When the goal is to image biological cells and tissues, at lower frequencies, ionic and interfacial polarization sensing takes place. However, at higher frequencies, these low-frequency barriers are bypassed, providing access to intracellular information and cytoplasmic changes \cite{Abasi_cellsens_acsmeas2022}. Therefore, the extreme sensitivity of mm-wave and THz signals to water content can be leveraged as a natural biomarker for non-invasive imaging of tissue malignancies \cite{Pfeiffer_uwavemag2019}. These signals further benefit from their shorter wavelengths, enabling a smaller pixel size and, thus, high-resolution imaging. Furthermore, the narrowband nature of resonators results in improved noise performance relative to wideband sensing. The combination of these distinctive characteristics has motivated researchers to develop high-resolution, high-contrast resonator-based mm-wave and THz near-field imaging systems.

Although maximizing the frequency of operation is desirable from a resolution standpoint, it poses challenges in delivering sufficient signal power to the imaging pixels and detecting their response efficiently without compromising sensitivity or resolution. 

\begin{figure}
    \centering
    \subfloat[]{\includegraphics[width=1.08in]{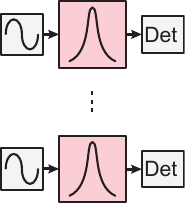}
    \label{fig:imgr_topol_1}}
    \subfloat[]{\includegraphics[width=1.49in]{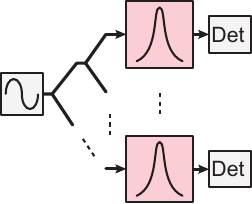}
    \label{fig:imgr_topol_2}}

    \subfloat[]{\includegraphics[width=2.34in]{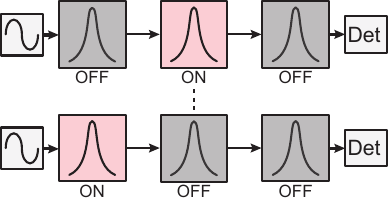}
    \label{fig:imgr_topol_3}}

    \caption{Resonator-based near-field imaging approaches. (\subref{fig:imgr_topol_1}) One detector and one source per pixel.(\subref{fig:imgr_topol_2}) One detector per pixel and one source per multiple pixels. (\subref{fig:imgr_topol_3}) One detector and one source per multiple pixels.}
    \label{fig:imager_apprch}
\end{figure}

The architectures proposed in \cite{ameri_vlsi2019,mitsunaka_jssc2016,tanaka_tcas2022}, as shown in Fig. \ref{fig:imager_apprch}(\subref{fig:imgr_topol_1}), were limited to using a single detector and a single signal source per imaging pixel. While these configurations achieve high sensitivity, the pixel size is determined not only by the resonator dimensions but also by the additional components of the source and detector. Therefore, the imaging resolution does not scale consistently with the frequency of operation.

A more effective approach was introduced in \cite{Grzyb_tmtt2017,hillger_jssc2018}, where pixels were coupled to, rather than directly connected to, the excitation source and detector. This allowed for separation among these components. This approach, as shown in Fig. \ref{fig:imager_apprch}(\subref{fig:imgr_topol_2}), enabled a single source to excite multiple pixels. However, each pixel still required its dedicated local detector, limiting the array scalability to a single dimension.

A key limitation of the discussed architecture is the inability to enable and disable imaging resonators selectively. This feature is essential for sharing a single source and a single detector among multiple pixels. By doing so, a two-dimensional array can be constructed that not only scales in both dimensions but also scales well with frequency, since each pixel consists solely of a resonator, whose size decreases with increasing frequency. This concept is depicted in Fig. \ref{fig:imager_apprch}(\subref{fig:imgr_topol_3}) and was demonstrated in \cite{ameri_isscc}, where animal tissue samples were imaged with high contrast at tens-of-microns resolution.

The reader is referred to \cite{ameri_jssc25} for a detailed discussion of the imager architecture and imaging experiments. However, the following provides a brief overview: a sub-THz signal is generated and split into two paths. The majority of the signal power is used as the LO signal for the detector. The remaining power is coupled to the RF path and is used for imaging. The RF path consists of a transmission line loaded with an array of imaging pixels. The sample shifts the phase characteristics of the pixels. These shifts are measured by sequentially enabling the pixels and performing phase detection by mixing the LO and RF signals. An image frame is constructed once all pixels are measured.

The most critical building block of this platform is the imaging pixel, which consists of SRRs equipped with active circuits to enable a tunable, boosted quality factor, with the ability to switch the pixel on and off. This paper develops a theoretical framework for analyzing various aspects of signal coupling, sensitivity, and noise in the imaging pixels. These analyses will lead to a set of design guidelines essential for designing an optimal two-dimensional near-field imaging system.

The paper is organized as follows. Section \ref{sec_srr} presents an equivalent circuit model for a lossy SRR upon which optimum coupling conditions are derived. Section \ref{sec_asrr} discusses sensing via ASRRs, maximum detection limits, and nonlinear effects. Section \ref{sec_noise_asrr} studies the impact of device white and flicker noise, supply noise, and input phase noise on the output phase noise. Finally, the pixel SNR is calculated in Section \ref{sec_snr}, providing design guidelines to maximize SNR while minimizing power consumption in a pixel array. Section \ref{sec_conclusion} concludes the paper.

\section{Split-Ring Resonator (SRR) Analysis}\label{sec_srr}

\begin{figure}[t]
    \centering
    \subfloat[]{\includegraphics[width=1.6in]{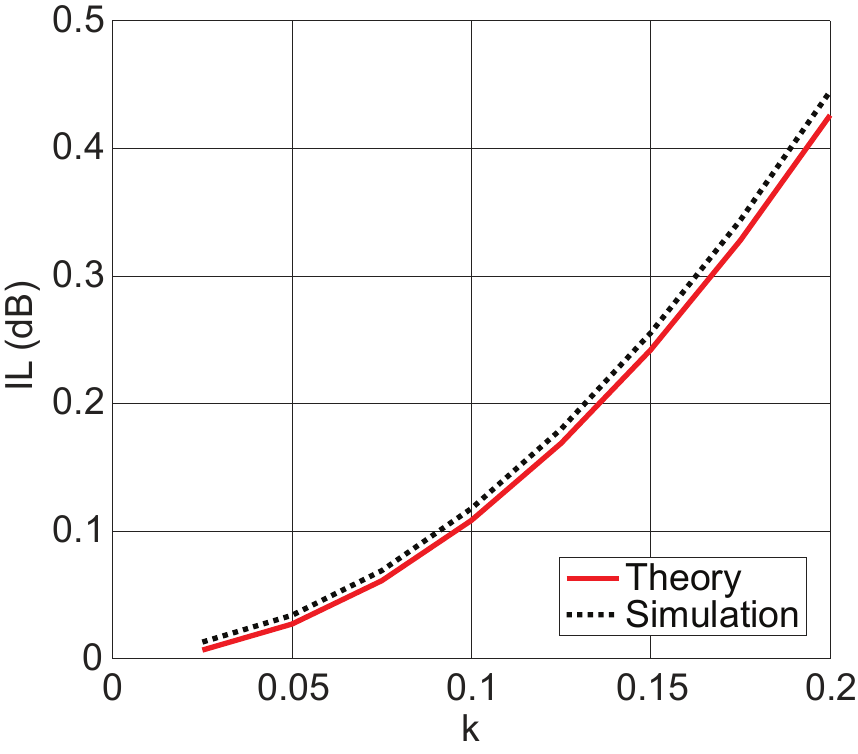}
    \label{fig:il_k}}
    \subfloat[]{\includegraphics[width=0.5in]{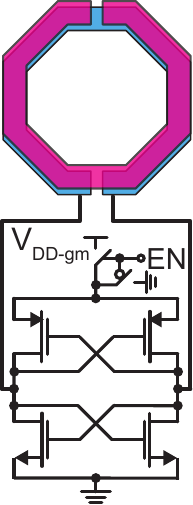}
    \label{fig:asrr}}
    \subfloat[]{\includegraphics[width=1.24in]{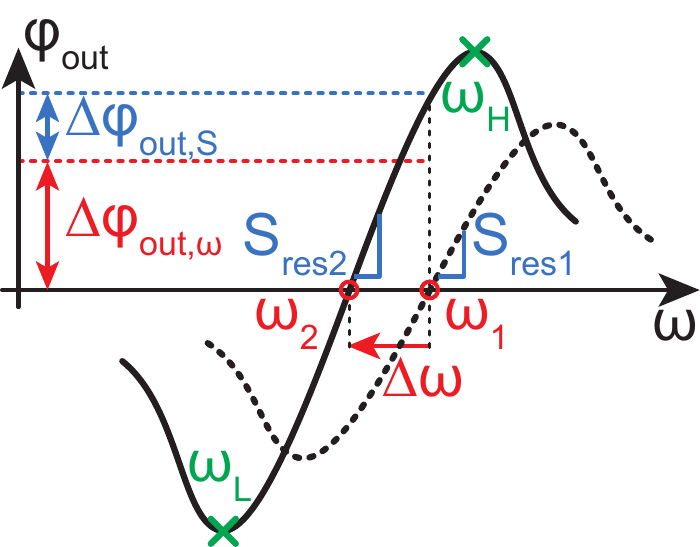}
    \label{fig:phi_out}}
    \caption{(\subref{fig:il_k}) IL as a function of coupling factor for a single resonator. (\subref{fig:asrr}) The schematic of an ASRR. (\subref{fig:phi_out}) The phase response of an ASRR and the relevant key parameters that change due to shifts in the resonance frequency and the quality factor.}
    \label{fig:il_asrr_res_curves}
\end{figure}

The resonator used in this work is a broadside-coupled SRR equipped with a positive feedback circuit, i.e., $-g_m$, collectively referred to as an ASRR, which serves as a single pixel in the imaging array. As shown in Fig. \ref{fig:il_asrr_res_curves}(\subref{fig:asrr}), the $-g_m$ block can be switched on and off to alternate between low-$Q$, $Q_{OFF}$, and high-$Q$, $Q_{ON}$, resonance modes, a feature essential to array functionality.  

\begin{figure}[t]
  \centering
  \includegraphics{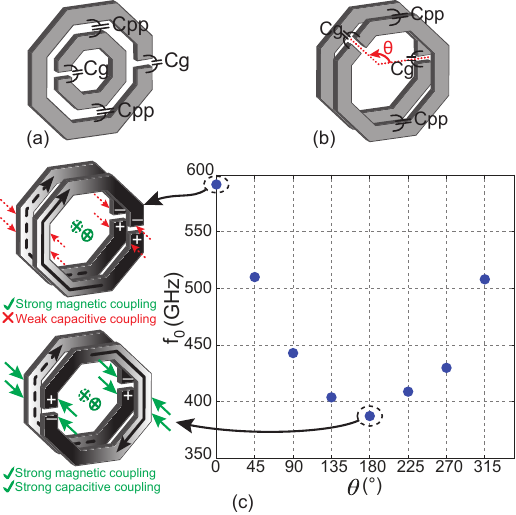}
  \caption{(a) Edge-coupled SRR, (b) Broadside-coupled SRR. (c) HFSS simulated resonance frequency as a function of $\theta$.}
  \label{fig:srr_ec_bc}
\end{figure}

In this section, the effects of $-g_m$ are set aside, and they will be reintroduced into our analyses in Section \ref{sec_asrr}. An SRR consists of two concentric rings with a gap opening on each ring. For understanding SRR EM behavior, the reader is referred to \cite{marques_physrevb2002}. There are multiple design choices for coupling the rings to each other and the placement of the gap on each ring. In imaging applications where the resonators must be as compact as possible, these layout choices, besides increasing the resonance frequency, can significantly reduce the resonator dimensions and thus increase the imaging resolution. In \cite{rmarques_srr_tant2003}, two possible coupling modes of the rings, namely edge coupling and broadside coupling, were studied. While both of these coupling modes can be easily implemented on chip, broadside coupling is preferred as it takes advantage of parallel plate capacitances between the two rings, which can be significantly higher compared to the parallel plate and fringing capacitance of the edge-coupled rings, as shown in Fig. \ref{fig:srr_ec_bc}(a). 

\begin{figure*}[t]
    \centering
    \subfloat[]{\includegraphics[width=1.3in]{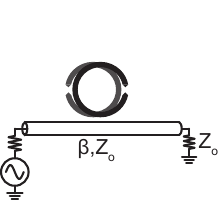}
    \label{fig:srr_cpl_tl}}
    \subfloat[]{\includegraphics[width=2in]{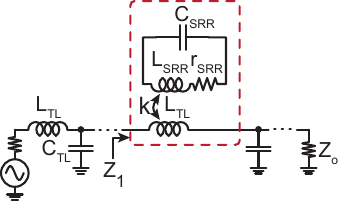}
    \label{fig:srr_cpl_tl_lc_1}}
    \subfloat[]{\includegraphics[width=1.8in]{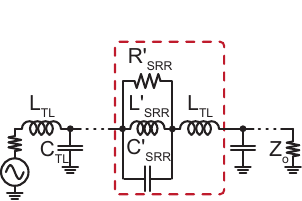}
    \label{fig:srr_cpl_tl_lc_2}}
    \subfloat[]{\includegraphics[width=1.5in]{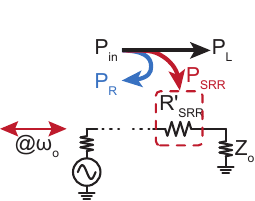}
    \label{fig:srr_at_res}}
    \caption{(\subref{fig:srr_cpl_tl}) An SRR coupled to a transmission line. (\subref{fig:srr_cpl_tl_lc_1}) The LC equivalent circuit of an SRR coupled to a transmission line. (\subref{fig:srr_cpl_tl_lc_2}) The equivalent SRR, modeled as a parallel resonator inserted in the transmission line at the point of coupling. (\subref{fig:srr_at_res}) The equivalent SRR at resonance.}
    \label{fig:srr_cpl}
\end{figure*}

\begin{figure}[t]
  \centering
  \subfloat[]{\includegraphics[width=1.5in]{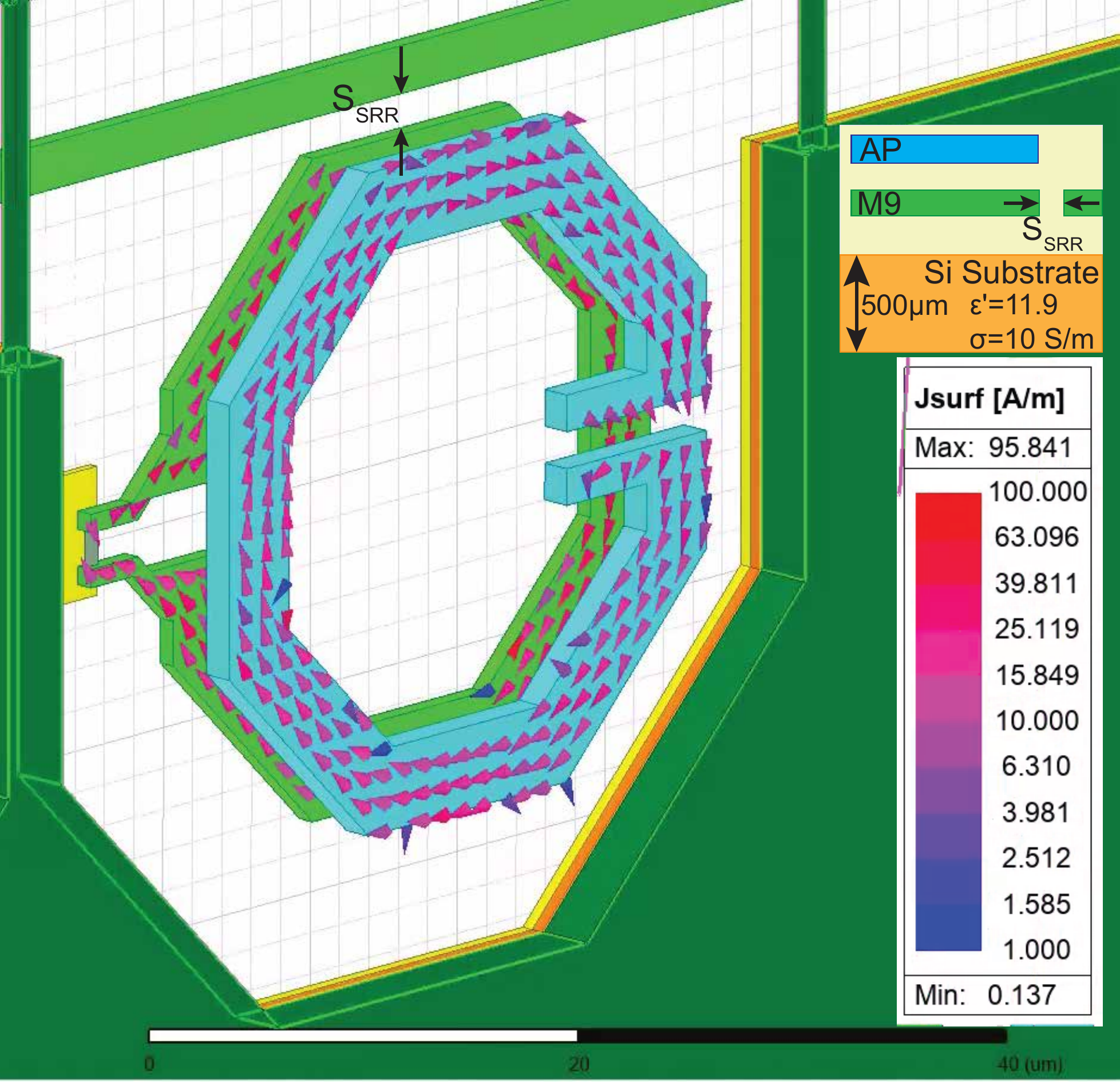}
  \label{fig:srr_em}}
  \subfloat[]{\includegraphics[width=1.8in]{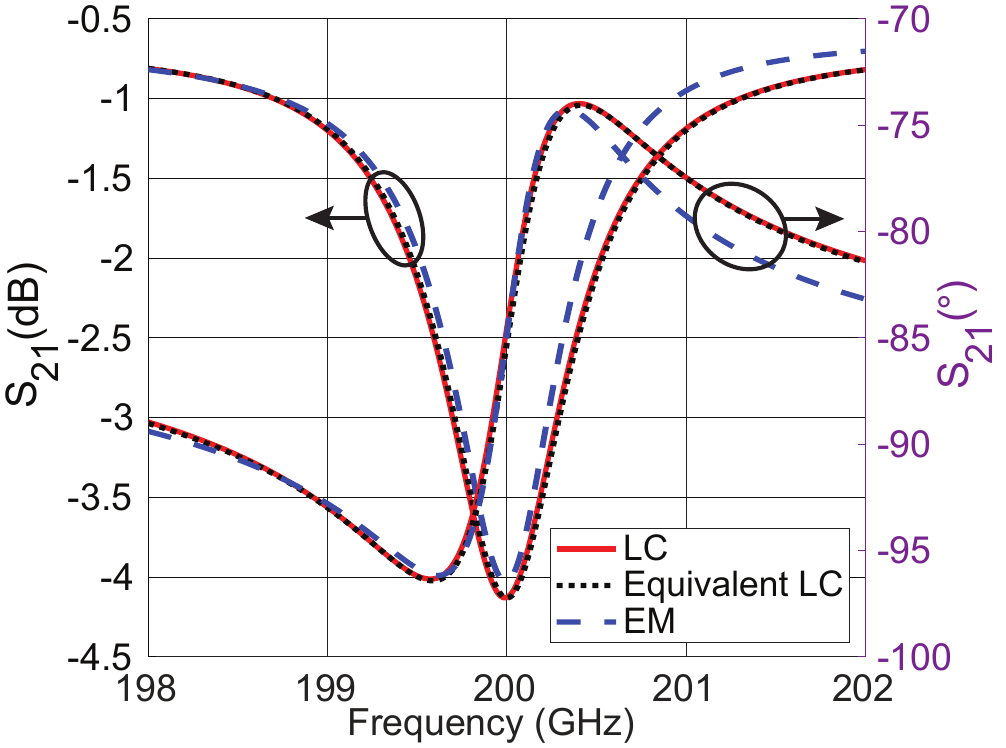}
  \label{fig:s21_lc_vs_em}}
  \caption{(\subref{fig:srr_em}) The HFSS EM model of an SRR. (\subref{fig:s21_lc_vs_em}) Simulated $S_{21}$ for the SRR LC, equivalent LC, and EM models.}
  \label{fig:s21_sim_em_model}
\end{figure}

In broadside-coupled resonators, the magnetic coupling is maximized for rings of the same size. At the lowest resonance frequency, the resonance currents must circulate in both rings in phase for the constructive summation of the magnetic fluxes and thus maximum inductance. In this mode, the positions of the gaps with respect to each other determine the extent to which the parallel plate capacitors between the rings contribute to the resonance. This is shown in Fig. \ref{fig:srr_ec_bc}(b), where $\theta$ denotes the relative angular position of the two gaps. $\theta=0^\circ$ means the gaps are aligned (symmetric), and $\theta=180^\circ$ means the gaps are on the opposite sides of the rings (anti-symmetric). In the symmetric case, the distributed parallel plate capacitors see no voltage difference across them and, therefore, do not affect the resonance. On the other hand, in the anti-symmetric case, a voltage difference exists between the distributed parallel plate capacitors, which increases the parasitic capacitance of the resonator and, in turn, reduces the resonance frequency. Fig. \ref{fig:srr_ec_bc}(c) shows the HFSS EM simulation results of two broadside-coupled resonators while sweeping $\theta$ from symmetric to anti-symmetric. The lowest unloaded resonance frequency for each $\theta$ is measured, which confirms that the anti-symmetric case achieves the lowest resonance frequency and is thus most suitable for the imaging applications.

\begin{figure*}[t]
\centering
\subfloat[]{\includegraphics[width=1.7in]{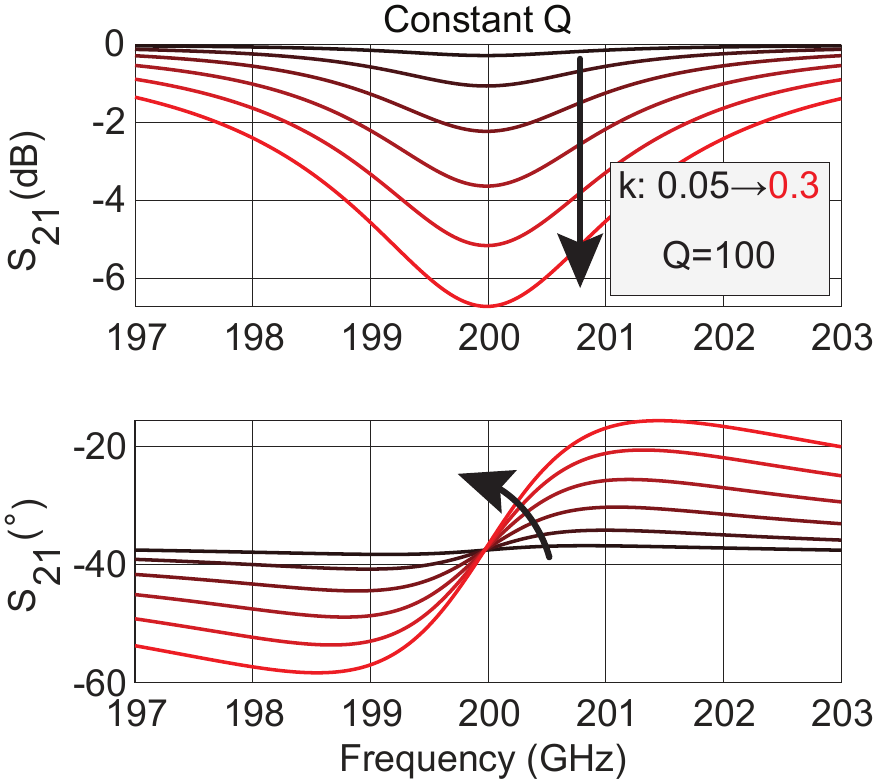}
\label{fig:s21_q_const}}
\subfloat[]{\includegraphics[width=1.7in]{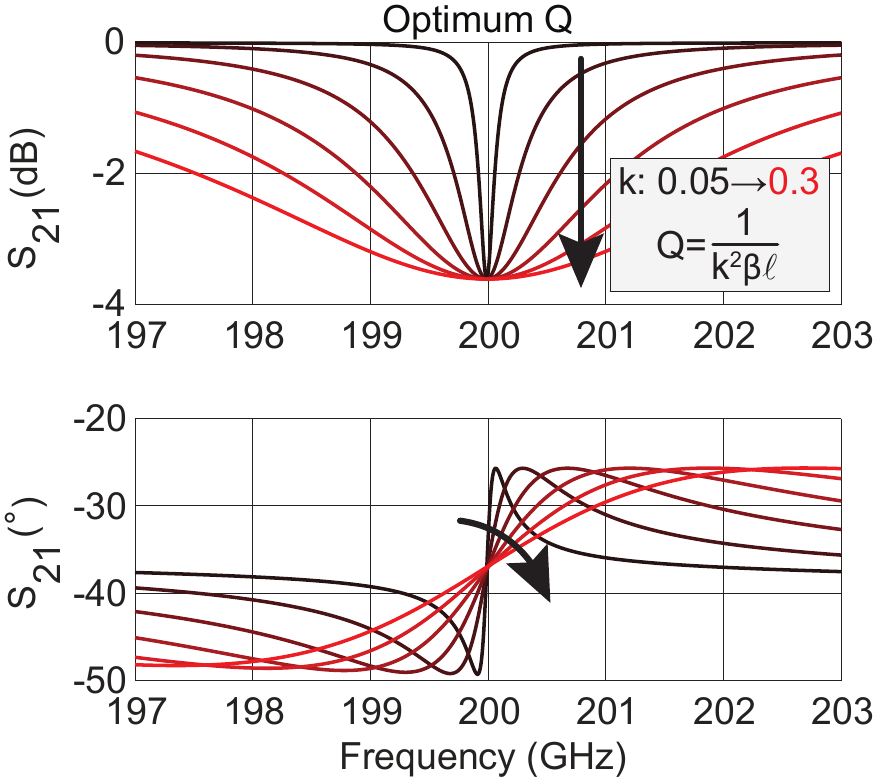}
\label{fig:s21_q_calc}}
\subfloat[]{\includegraphics[width=1.7in]{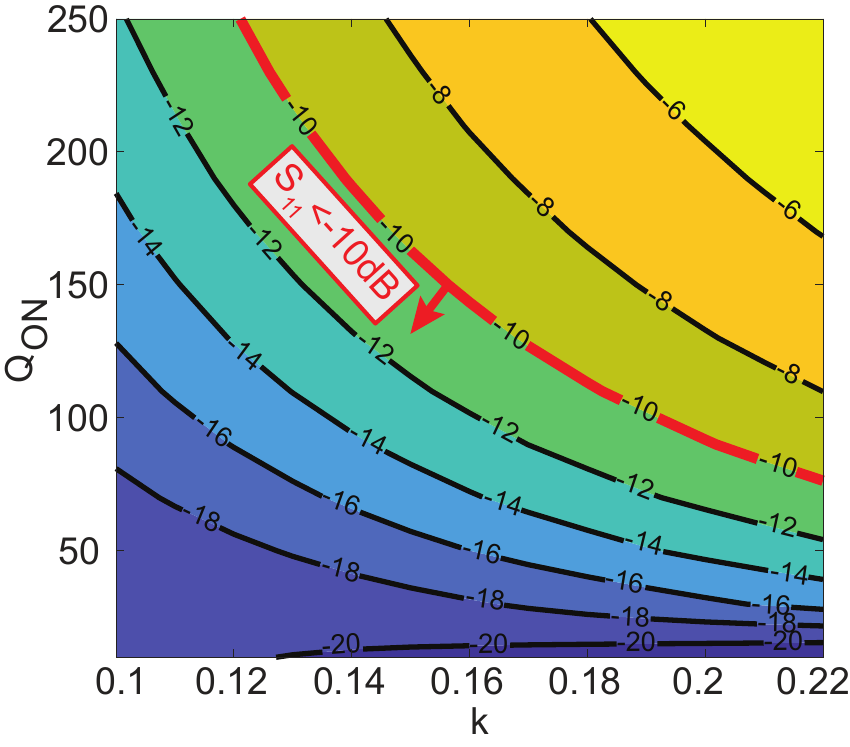}
\label{fig:s11_cntrs}}
\subfloat[]{\includegraphics[width=1.85in]{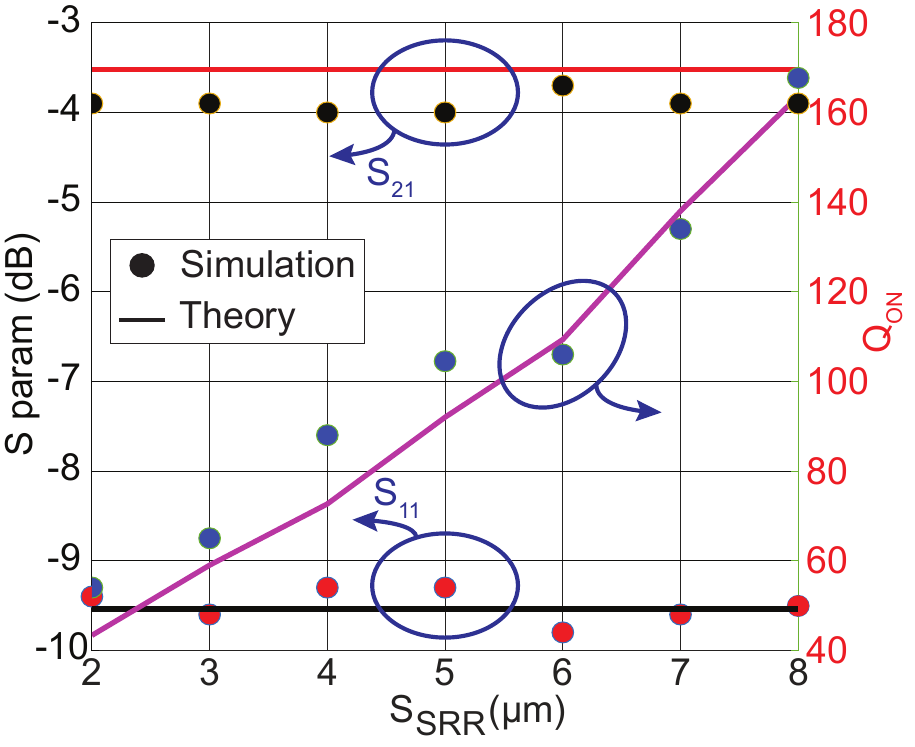}
\label{fig:sp_q_calc_sim}}
\caption{The magnitude and phase of $S_{21}$ while sweeping $k$ for (\subref{fig:s21_q_const}) constant $Q$, and (\subref{fig:s21_q_calc}) calculated $Q$ from (\ref{eq_kq_srr}), where $\beta l=0.25$. (\subref{fig:s11_cntrs}) The contours of $S_{11}$ for various $(k,Q_{ON})$ pairs and the boundary where $S_{11}<-10dB$. (\subref{fig:sp_q_calc_sim}) $S_{11}$, $S_{21}$, and $Q_{ON}$ of EM simulated SRR compared to the calculated values as a function of SRR spacing to the host transmission line, and therefore $k$.}
\label{fig:sp_sims}
\end{figure*}

\subsection{Equivalent Circuit Models}
The SRR coupled to a lossless transmission line, with a phase constant of $\beta$ and a characteristic impedance of $Z_0$, as shown in Fig. \ref{fig:srr_cpl}(\subref{fig:srr_cpl_tl}) can be modeled using a distributed LC model for the transmission line and a lumped LC model for the resonator as illustrated in Fig. \ref{fig:srr_cpl}(\subref{fig:srr_cpl_tl_lc_1}). Here, $L_{TL}$ and $C_{TL}$ represent the distributed inductance and capacitance values of the transmission line section of length $l$. As such, $\beta=\omega\sqrt{L_{TL}C_{TL}}/l$ and $Z_o=\sqrt{L_{TL}/C_{TL}}$. The losses of the transmission line can be neglected as $l$ is small (on the order of the SRR length, which is $\sim$30-40 $\mu m$). The SRR is modeled via $L_{SRR}$ and $C_{SRR}$, where the losses are modeled with $r_{SRR}=\omega_0L_{SRR}/Q$. The capacitive couplings between the SRR and the transmission line are negligible compared to the magnetic couplings, which is denoted by $k$. The impedance, $Z_1$, of the combination of SRR coupled to the transmission line, assuming that it is terminated by $Z_o$, is given by

\begin{equation}
    Z_1(s)=sL_{TL}-\frac{s^2M^2}{r_{SRR}+sL_{SRR}+\frac{1}{sC_{SRR}}}
    \label{eq_z1s}
\end{equation}
where $M=k\sqrt{L_{TL}L_{SRR}}$. Substituting $s=j\omega$ in (\ref{eq_z1s}) and manipulating the RHS yields
\begin{equation}
    Z_1(j\omega)=j\omega L_{TL}+\frac{1}{\frac{r_{SRR}}{\omega^2M^2}+j\omega \frac{L_{SRR}}{\omega^2M^2}+\frac{1}{j\omega (\omega^2M^2C_{SRR})}}.
    \label{eq_z1w}
\end{equation}

The impedance given by (\ref{eq_z1w}), around the resonance frequency $\omega _0$, can be interpreted as $L_{TL}$ in series with a parallel equivalent $RLC$ resonator with component values given by

\begin{subequations}
    \begin{align}
        &R_{SRR}^{'}=({\omega_0^2M^2})/{r_{SRR}}\label{eq_srr_rp}\\
        &L_{SRR}^{'}={\omega_0}^2M^2C_{SRR}\\
        &C_{SRR}^{'}={L_{SRR}}/({{\omega_0}^2M^2})
    \end{align}
    \label{eq_srr}
\end{subequations}
where $\omega_0=1/{\sqrt{L_{SRR}C_{SRR}}}=1/{\sqrt{{L^{'}}_{SRR}{C^{'}}_{SRR}}}$ is the SRR resonance frequency. This equivalent model is depicted in Fig. \ref{fig:srr_cpl}(\subref{fig:srr_cpl_tl_lc_2}). These calculations supplement the results presented in \cite{fmartin_appphylet2003,baena_srr_tmtt2005,shang_srr_tmtt2013} using other methods for lossless resonators. Including the loss has numerous implications in the design for a proper input match, as well as sensitivity and noise analyses. Around $\omega_0$, the circuit reduces to Fig. \ref{fig:srr_cpl}(\subref{fig:srr_at_res}), which will be used to derive the optimum coupling conditions subsequently.

It is crucial to demonstrate that the developed LC models closely mimic the response of a 3D EM model, which is considered the gold standard. Figure \ref{fig:s21_sim_em_model}(\subref{fig:srr_em}) illustrates the 3D model of a broadside-coupled SRR coupled to a 50$\Omega$ transmission line simulated in HFSS. The figure also shows the surface current density at resonance and a cross-sectional layout view, where the bottom ring and the transmission line are implemented using M9, and the top ring is on the AP layer. The simulated magnitude and phase of $S_{21}$ for the LC model (Fig. \ref{fig:srr_cpl}(\subref{fig:srr_cpl_tl_lc_1})), the \textit{equivalent} LC model (Fig. \ref{fig:srr_cpl}(\subref{fig:srr_cpl_tl_lc_2})) and the EM models are plotted in Fig. \ref{fig:s21_sim_em_model}(\subref{fig:s21_lc_vs_em}). The RLCk parameters for the LC model were first derived by curve-fitting its response to that of the EM model. These values were then used to find the component values in the equivalent LC model, based on (\ref{eq_srr}). The close agreement among the simulation results of these models demonstrates that the LC model can be effectively used in lieu of the EM model, which drastically streamlines the subsequent analyses throughout this paper.

\subsection{Coupling Effects and Optimum Coupling}\label{sec_coupling}

At first glance, a stronger coupling between the SRR and the transmission line seems preferable. Fig. \ref{fig:sp_sims}(\subref{fig:s21_q_const}) shows the magnitude and phase of the forward transmission, $S_{21}$, for an SRR tuned at 200GHz while changing the coupling coefficient, $k$, for a fixed quality factor of 100. For a stronger coupling, the depth of the notch in the magnitude and the slope of the phase of $S_{21}$ increase, both of which imply a higher sensitivity to a shift in resonance characteristics, and thus preferable for sensing. To clarify, although the unloaded resonator $Q$ at the targeted frequencies never exceeds $10-20$, as will be seen, once loaded with an $-g_m$ block, the $Q$ can be boosted to much higher values. As such, in the following discussions, we will assume that the $Q_{ON}$, the boosted $Q$ by $-g_m$ can greatly exceed its typical unloaded values. On the other hand, $Q_{OFF}$, the quality factor when the $-g_m$ is disabled, is the same as the unloaded resonator $Q$.

An SRR, by loading a section of its host transmission line, both reflects and absorbs the incident power, as shown in Fig. \ref{fig:srr_cpl}(\subref{fig:srr_at_res}). $P_R$ and $P_{SRR}$ denote the reflected and absorbed powers, and $P_L$ is the power delivered to the load. To provide sufficient matching to the source, an $S_{11}=1/3$ can be obtained when $R_{SRR}^{'}=Z_o$\footnote{To be exact, $Z_o/2\leq R_{SRR}^{'}\leq Z_o$ results in $|S_{11}|\leq1/3$. However, the maximum allowed value is chosen to achieve the maximum quality factor.}. Using the expressions in (\ref{eq_srr}a), one arrives at the following insightful relationship between the coupling and quality factors of the SRR
\begin{equation}
    \beta lk^2Q_{ON}=1.
    \label{eq_kq_srr}
\end{equation}

For optimal coupling, that is, for input matching, either the $Q_{ON}$ or $k$ is chosen, allowing the other parameter to be determined according to (\ref{eq_kq_srr}). Fig. \ref{fig:sp_sims}(\subref{fig:s21_q_calc}) depicts the magnitude and phase of $S_{21}$ for the 200GHz SRR, while sweeping $k$ and using the optimum $Q_{ON}$, calculated from (\ref{eq_kq_srr}), for each $k$. One can see that the magnitude of $S_{21}$ at resonance stays constant by choosing the proper quality factor for a coupling coefficient. Another way of looking at this is by performing a two-dimensional sweep of $(Q_{ON},k)$ pairs on the circuit shown in Fig. \ref{fig:srr_cpl}(\subref{fig:srr_cpl_tl_lc_1}). Fig. \ref{fig:sp_sims}(\subref{fig:s11_cntrs}) illustrates the contours of $S_{11}$ as a function of $k$ and $Q_{ON}$. The highlighted line specifies the region where $S_{11}=-10dB$. It is worth noting that $k$ is determined by the design geometry and is therefore not tunable after fabrication. $Q_{ON}$ is, however, tunable as it is set by the variable transconductance that activates the resonators.

Equation (\ref{eq_kq_srr}) plays a crucial role in the subsequent analyses and has been verified through simulations. In the EM model presented earlier, $k$ is varied by changing the space between the resonator and the transmission line, denoted by the variable $S_{SRR}$, in Fig. \ref{fig:sp_sims}(\subref{fig:sp_q_calc_sim}). The value of $k$ for each $S_{SRR}$ was determined through curve-fitting of the LC model response to that of the EM model. $\beta$ can also be derived from the EM model by characterizing the host transmission line, and $l$ is the length of the segment of the line that interacts with the resonator. Then, the requisite $Q_{ON}$ was calculated from (\ref{eq_kq_srr}) and was realized using an ideal negative resistance placed in parallel with the SRR. The calculated and simulated $Q_{ON}$ values to achieve optimum coupling are plotted against each other in Fig. \ref{fig:sp_sims}(\subref{fig:sp_q_calc_sim}). More importantly, this figure also contains the resulting return loss and forward transmission for different coupling configurations. It can be seen that an $S_{11}<-9.25dB$ and $S_{21}>-4dB$ are consistently achieved as predicted by the theory, which confirms the validity of (\ref{eq_kq_srr}) and the utility of the LC model.

The foregoing discussion limits the $(Q_{ON},k)$ design space based on a desired $S_{11}$. However, in coupling resonators to a transmission line, there is an upper limit dictated by the geometry and layout rules in addition to the total insertion loss due to the coupling of multiple pixels to the transmission line. As the $k$ value is mainly determined by the proximity of the SRR to the transmission line, its value can barely exceed 0.25 based on EM simulations. The $S_{21}$ from the equivalent LC model can be calculated from 

\begin{equation}
    S_{21}=\frac{2Z_0}{Z^{'} + 2Z_0}
    \label{eq_s21}
\end{equation}
where $Z^{'}$ is the impedance of the equivalent parallel $R^{'}L^{'}C^{'}$ resonator. In an array of $N$ passive resonators, based on (\ref{eq_s21}) and (\ref{eq_srr_rp}), the total Insertion Loss (IL) at resonance can be written as

\begin{equation}
    IL(N)=N\frac{\omega_0k^2Q_{OFF}L_{TL}}{\omega_0k^2Q_{OFF}L_{TL}+2Z_0}
    \label{eq_il}
\end{equation}

To achieve a specific insertion loss, a maximum coupling coefficient, $k_{M\!\!AX}$, can be calculated from (\ref{eq_il}), as every parameter is known except for $k$. This sets an upper bound on the coupling factor, which in turn, according to (\ref{eq_kq_srr}), sets a lower bound on $Q_{ON}$ as

\begin{equation}
    Q_{ON,min}=\frac{1}{\beta lk_{MAX}^2}
    \label{eq_qon_min}
\end{equation}

These discussions will become crucial when designing for a target SNR in a single pixel, and will be revisited in Section \ref{sec_snr}. Fig. \ref{fig:il_asrr_res_curves}(\subref{fig:il_k}) depicts the simulated insertion loss due to a single resonator for different coupling factors and compares it with $IL(1)$ given by (\ref{eq_il}).

\subsection{Signal Detection}\label{sec_sig_det}

\begin{figure*}[t]
    \centering
    \subfloat[]{\includegraphics[width=3.6in]{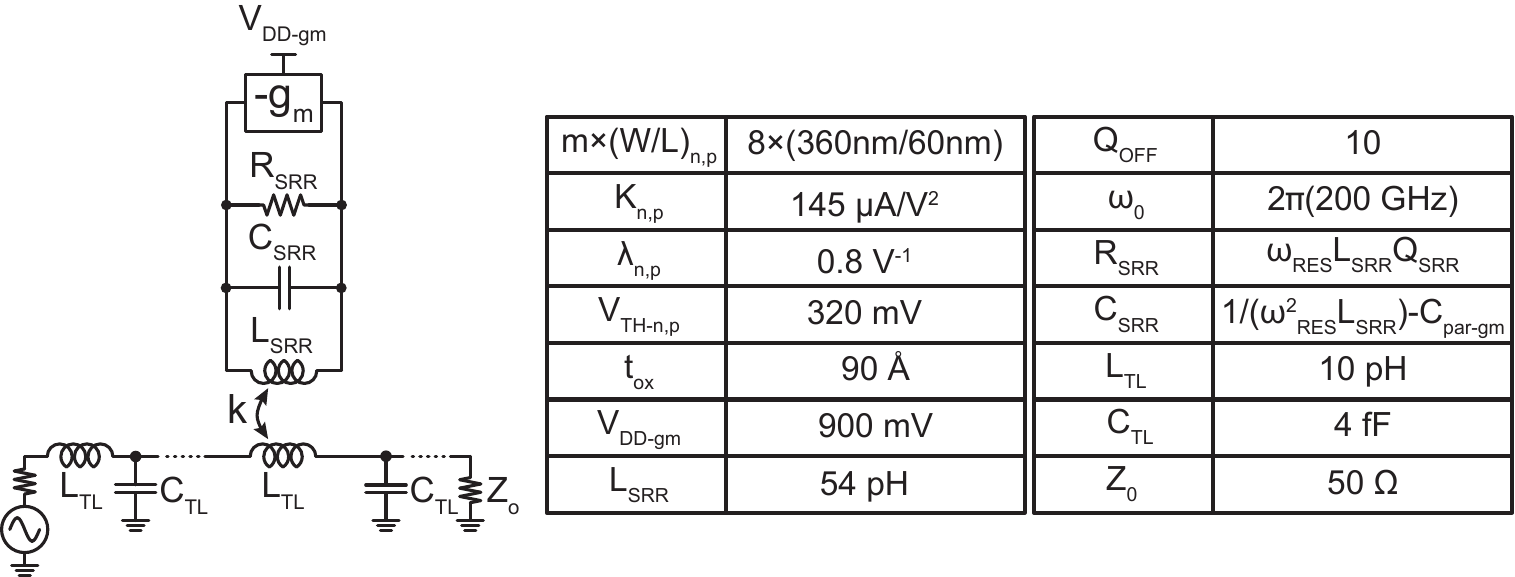}
    \label{fig:asrr_lc_eq}}    
    \subfloat[]{\includegraphics[width=1.7in]{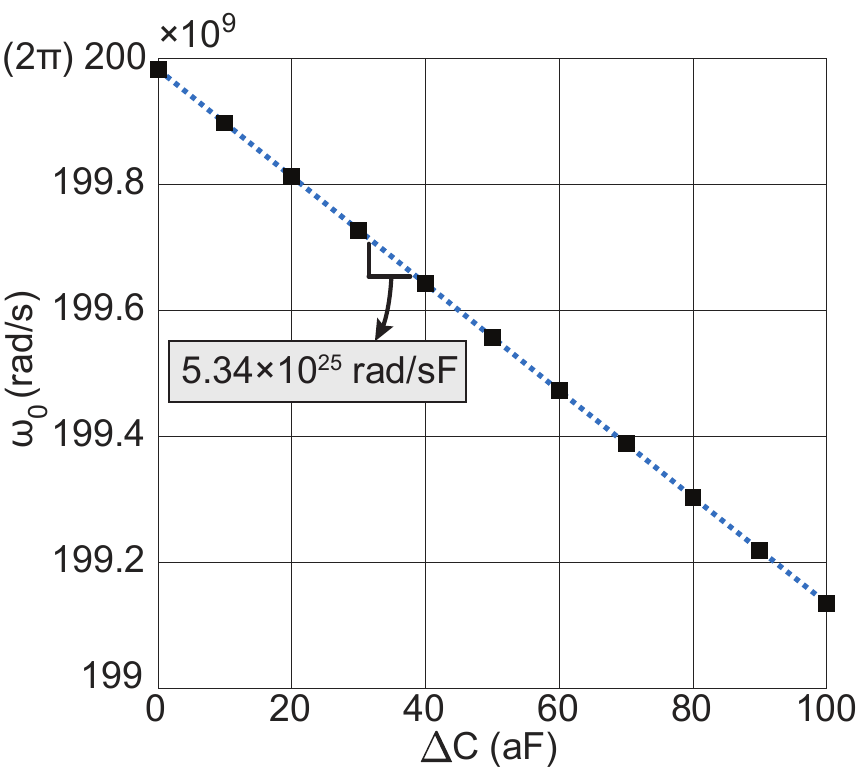}
    \label{fig:dwres_dc}}
    \subfloat[]{\includegraphics[width=1.7in]{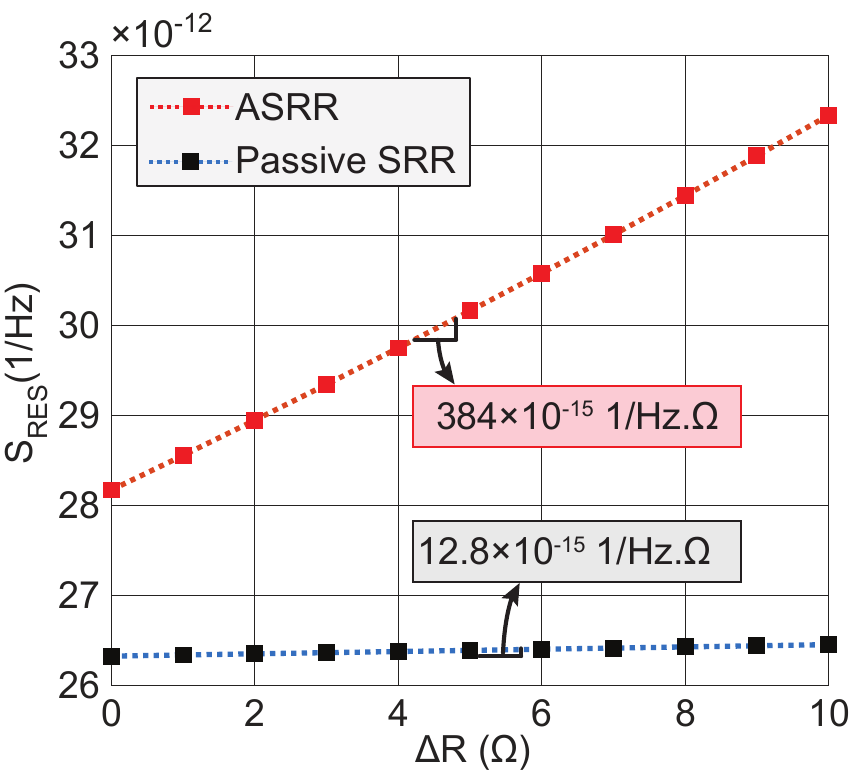}
    \label{fig:dsres_dr}}

    \caption{(\subref{fig:asrr_lc_eq}) Equivalent LC model and the circuit parameters for a 200GHz ASRR used for analysis and simulations. (\subref{fig:dwres_dc}) $\omega_{0}$ as a function of $\Delta C$. (\subref{fig:dsres_dr}) $S_{RES}$ as a function of $\Delta R$.}
    \label{fig:asrr_lc_sens_sim}
\end{figure*}

The foregoing discussion indicates that a parallel LC resonance response is expected from an SRR, characterized by a notch in the amplitude and a high-slope region in the phase of the output signal. This provides the opportunity for both power and phase detection. Power detection, due to its wideband operation and compactness, is a widely used technique in high-frequency sensing and spectroscopy systems, such as in \cite{hillger_jssc2018,wang_gasspec_jssc2021,laemmle_sens_tmtt2013}. However, the high phase noise of sub-THz VCOs can severely compromise the detection SNR. Alternatively, phase detection provides immunity to the VCO phase noise. As will be shown in Section \ref{sec_inp_pnoise}, the pixel input and output close-in phase noise are identical, and thus the phase noise, as a correlated signal, gets suppressed by phase detection. Since each pixel operates in a very narrow band around its resonance frequency, a quadrature phase relationship between the $LO$ and $RF$ signals can be established by introducing an adequate phase shift in one of these signal paths. Consequently, the output of the phase detector can be expressed as 

\begin{align}
    V_{out,DET}&=v_{LO}cos(\omega t+\varphi_{n})\cdot v_{RF}cos(\omega t+\frac{\pi}{2}+\varphi+\varphi_{n})\nonumber\\
    &=\frac{1}{2}v_{LO}v_{RF}sin(\varphi)
\end{align}
where $v_{RF}$ and $v_{LO}$ are amplitudes of the $RF$ and $LO$ signals, respectively, $\varphi_n$ is the VCO phase noise, $\varphi$ is the extra phase induced by the sample in the $RF$ path, which is the desired signal, and the second harmonic term is assumed to be filtered out after mixing. The amplitude noise of both $v_{RF}$ and $v_{LO}$ is highly attenuated by a small transfer function to the output due to $sin(\varphi)\ll1$ for small $\varphi$. As a result, for the noise analysis, only the contribution of different noise sources to $\varphi$, which corrupts our desired signal, will be studied.

Defining $\varphi_{out}$ as the phase of the output signal, the quantity $d\varphi_{out}/d\omega$ is of interest. One can see that a change in the resonance frequency due to sensing a sample gets translated to a phase difference scaled by a factor of $d\varphi_{out}/d\omega$. This quantity has been derived in Appendix \ref{appa} (\ref{eq_dphi_dw_2}), which for a $-10dB$ return loss, i.e. $R'=Z_0$, simplifies to
\begin{equation}
    \frac{d\varphi_{out}}{d\omega}=\frac{2R'}{3\omega_0^2L'}=\frac{2}{3}\frac{Q_{ON}}{\omega_0}=\frac{1}{3}(\frac{d\varphi}{d\omega})_{SRR}
    \label{eq_dphi_dw_mtch}
\end{equation}

Equation (\ref{eq_dphi_dw_mtch}) implies that only one-third of the sensitivity offered by the SRR appears at the output. This loss of performance is expected and ascribed to imperfect coupling between the resonator and the transmission line. Alternatively, for an optimally coupled SRR, one can define an effective quality factor, $Q_{out}$, based on the output phase response as

\begin{equation}
    Q_{out}=\frac{d\varphi_{out}}{d\omega}\frac{\omega_0}{2}=\frac{Q_{ON}}{3}
    \label{eq_qout}
\end{equation}


\section{Active SRR (ASRR) Analysis}\label{sec_asrr}

The quality factor numbers discussed above can never be achieved in current CMOS technologies. The $Q$ for a well-designed sensing resonator at 200 GHz is typically between 10 and 20, but this value is severely degraded once the resonator is exposed to water-rich, lossy biological samples. However, one can leverage the fact that these resonators are electrically isolated and equip them with active devices in positive feedback, to compensate for some of the losses of the resonators, thereby boosting their $Q$. From a circuit's perspective, this is equivalent to introducing a negative resistance, $-1/g_m$, to the SRR, essentially turning a passive resonator into an active one. An ASRR, as depicted in Fig. \ref{fig:il_asrr_res_curves}(\subref{fig:asrr}) has two key features: 1-a tunable quality factor, $Q_{ON}$, by changing the $-g_m$ supply voltage, $V_{DD-g_m}$, and 2-it can be enabled and disabled, by turning the $-g_m$ block on and off with a supply switch. These features play vital roles in setting the pixel sensitivity, array functionality, calibration, and noise suppression.

In an ASRR, $R_{ASRR}$, the equivalent boosted resistance, is given by
\begin{equation}
    R_{ASRR}=\frac{R_{SRR}}{1-g_mR_{SRR}}
    \label{eq_r_asrr}
\end{equation}
where $R_{SRR}=\omega_0 L_{SRR}Q_{OFF}$. The boosted quality factor of ASRR, $Q_{ON}$ can, therefore, be written as 

\begin{equation}
    Q_{ON}=\frac{R_{ASRR}}{\omega_0L_{SRR}}=\frac{Q_{OFF}}{1-g_mR_{SRR}}.
    \label{eq_qon}
\end{equation}

One must also note that the SRR, in addition to its intrinsic capacitance, is loaded with the total $-g_m$ parasitic capacitors, given by $C_{gm}=(C_{gs,n}+C_{gs,p}+C_{db,n}+C_{db,p})/2+(C_{gd,n}+C_{gd,p})\times2$. Therefore, $C_{ASRR}=C_{SRR}+C_{gm}$.

The introduction of the sample to an ASRR changes both $C_{SRR}$ and $R_{SRR}$. These changes can be expressed as $\Delta C=f^{'}(\Delta\Re({\varepsilon}))$ and $\Delta R=f^{''}(\Delta\Im({\varepsilon}))$, as studied in \cite{Ferrier_loc,Elhadidy_sens_tcas2015,chien_reactsens_jssc2016}, where $\varepsilon=\Re({\varepsilon})+j\Im({\varepsilon})$ is the complex permittivity of the sample. In the case of passive SRR, $\Delta C$ and $\Delta R$ directly alter the resonance frequency and quality factor, respectively. However, in an ASRR, although $\Delta C$ still directly changes the resonance frequency, the effect of $\Delta R$ is amplified by the Q-boosting factor squared, using (\ref{eq_r_asrr}) and (\ref{eq_qon}) one arrives at

\begin{align}
    \Delta R_{ASRR}&=\frac{dR_{ASRR}}{dR_{SRR}}\Delta R=\left(\frac{Q_{ON}}{Q_{OFF}}\right)^2\Delta R.
    \label{eq_del_rasrr}
\end{align}

This amplification factor plays a crucial role in high-contrast imaging, particularly in studying tissue hydration levels, where the goal is to measure the concentration of water, which exhibits high losses, i.e., $\Delta R\gg1$.

The addition of the active circuit to the SRR comes with the cost of introducing multiple sources of noise, which will be treated in the following discussions. All deviations from the SRR original operating point, whether desirable due to the sample or undesirable due to noise, can be decomposed into changes in the resonance frequency and the quality factor, as shown in Fig. \ref{fig:il_asrr_res_curves}(\subref{fig:phi_out}). This decomposition greatly simplifies the study of signal and noise behaviors in a pixel. The slope of the phase response around the resonance is defined as $S_{RES}\triangleq d\varphi_{out}/d\omega$. Thus, the total changes in the output phase, $\Delta\varphi_{out}$, can be attributed to two sources as

\begin{subequations}
\begin{align}
    &\Delta\varphi_{out,\omega}=-S_{RES}\Delta \omega_{0}\label{eq_delphi_out_w}\\
    &\Delta\varphi_{out,S}=-\Delta S_{RES}\Delta \omega_{0}\label{eq_delphi_out_s}
\end{align}
    \label{eq_delphi_out}
\end{subequations}
where $\Delta \omega_{0}$ and $\Delta S_{RES}$ are the changes in the resonance frequency and the changes in the derivative of $\varphi_{out}$ around the resonance frequency, respectively. These quantities in the presence of a sample can be computed as follows

\begin{equation}
    \Delta \omega_{0}=-\frac{\Delta C}{2C_{ASRR}}\omega_{0}
    \label{eq_dwres_dc}
\end{equation}
and $\Delta S_{RES}$ can be calculated from (\ref{eq_dphi_dwdr_2}), under optimum coupling, using (\ref{eq_kq_srr}) and (\ref{eq_del_rasrr}) it becomes

\begin{align}
    \Delta S_{RES}&=\frac{10}{9}\frac{1}{\beta l\omega_0Z_0}\frac{L_{TL}}{L_{SRR}}\Delta R_{ASRR}\nonumber\\
    &=\frac{10}{9}C_{ASRR}\left(\frac{Q_{ON}}{Q_{OFF}}\right)^2\Delta R
    \label{eq_dsres_dr}
\end{align}

Utilizing (\ref{eq_dphi_dw_mtch}) and substituting (\ref{eq_dwres_dc}) and (\ref{eq_dsres_dr}) in (\ref{eq_delphi_out_w}) and (\ref{eq_delphi_out_s}) results in

\begin{subequations}
\begin{align}
    &\Delta\varphi_{out,\omega}=\frac{Q_{ON}}{3}\frac{\Delta C}{C_{ASRR}}\label{eq_delphi_out_w_2}\\
    &\Delta\varphi_{out,S}=\frac{5}{9}\left(\frac{Q_{ON}}{Q_{OFF}}\right)^2\omega_{0}\Delta R\Delta C\label{eq_delphi_out_s_2}
\end{align}
    \label{eq_delphi_out_2}
\end{subequations}
which can be used for deriving the output signal due to the changes caused by the sample. An equivalent LC model for an ASRR was devised, as illustrated in Fig. \ref{fig:asrr_lc_sens_sim}(\subref{fig:asrr_lc_eq}). In this model, the $-g_m$ was implemented using both a square-law model and the standard 28nm PDK device models. Interestingly, as will be seen in the following sections, the square-law model proved to be accurate in all signal flow, noise, and nonlinearities analyses, when compared to PDK models.

The circuit in Fig. \ref{fig:asrr_lc_sens_sim}(\subref{fig:asrr_lc_eq}) with the component values summarized in the table was used to verify these derivations. $C_{SRR}$ and $R_{SRR}$ were varied while measuring $\omega_{0}$ and $S_{RES}$. Based on (\ref{eq_dwres_dc}), $\Delta \omega_{0}/\Delta C_{SRR}=-5.35\times 10^{25}$ (rad/s.F), which matches the slope of the curve shown in Fig. \ref{fig:asrr_lc_sens_sim}(\subref{fig:dwres_dc}). 

For simulating the effect of $\Delta R$, two cases of passive SRR and ASRR with equal quality factors, $Q=54$, were considered. For the passive SRR, a high-Q resonator was simply used, and for the ASRR, a resonator with $Q_{OFF}=10$ was employed, whose Q was boosted to 54, i.e., $Q_{ON}/Q_{OFF}=54/10$. $\Delta S_{RES}/\Delta R$ without the Q-boosting factor for a passive SRR is $10/9C_{SRR}=13\times 10^{-15}$ (1/Hz.$\Omega$). With the Q-boosting effect, $\Delta S_{RES}/\Delta R$ is given by (\ref{eq_dsres_dr}) and evaluates to $380\times 10^{-15}$ (1/Hz.$\Omega$). These calculations agree well with the slope of the lines shown in Fig. \ref{fig:asrr_lc_sens_sim}(\subref{fig:dsres_dr}) for each resonator.  

\subsection{Maximum Detection Limit}

The $Q$ enhancement in ASRRs creates a tradeoff on the maximum detection limit as well. The pixel is considered saturated if the shift in the resonance frequency is large enough so that the measurement frequency falls outside of the region where $S_{RES}>0$. This usable detection range can be estimated by finding the frequencies where the $\angle{S_{21}}$ reaches its extrema, as denoted by $\omega_H$ and $\omega_L$ in Fig. \ref{fig:il_asrr_res_curves}(\subref{fig:phi_out}). That is, frequencies where ${d\varphi_{out}}/{d\omega}=0$. In Appendix \ref{appa}, an approximate expression for ${d\varphi_{out}}/{d\omega}$ about the resonance frequency is derived, which cannot be used here, as the focus is more on the behavior away from the resonance. Nevertheless, based on (\ref{eq_dphi_dw_1}), and the fact that ${d\varphi_{out}}/{d\omega}\approx{d\Im(Z')}/{d\omega}$, according to (\ref{eq_rlc_imp}), the extrema can be found from

\begin{equation}
    \frac{d}{d\omega}\left[\frac{\omega L'(1-\frac{\omega^2}{\omega_0^2})}{(1-\frac{\omega^2}{\omega_0^2})^2+\frac{1}{Q_{ON}^2}(\frac{\omega^2}{\omega_0^2})}\right]=0
\end{equation}
which, after carrying out the differentiation, results in

\begin{equation}
    (1-\frac{\omega^2}{\omega_0^2})^2-\frac{1}{Q_{ON}^2}(\frac{\omega^2}{\omega_0^2})=0
\end{equation}
which has two positive solutions

\begin{align}
    \omega_{L,H}&=\omega_0\sqrt{\frac{2+\frac{1}{Q_{ON}^2}\pm\frac{1}{Q_{ON}}\sqrt{4+\frac{1}{Q_{ON}^2}}}{2}}
    \label{eq_srr_bw}\approx\omega_0\sqrt{1\pm\frac{1}{Q_{ON}}}.
\end{align}

\begin{figure*}[t]
    \centering
    \subfloat[]{\includegraphics[width=1.7in]{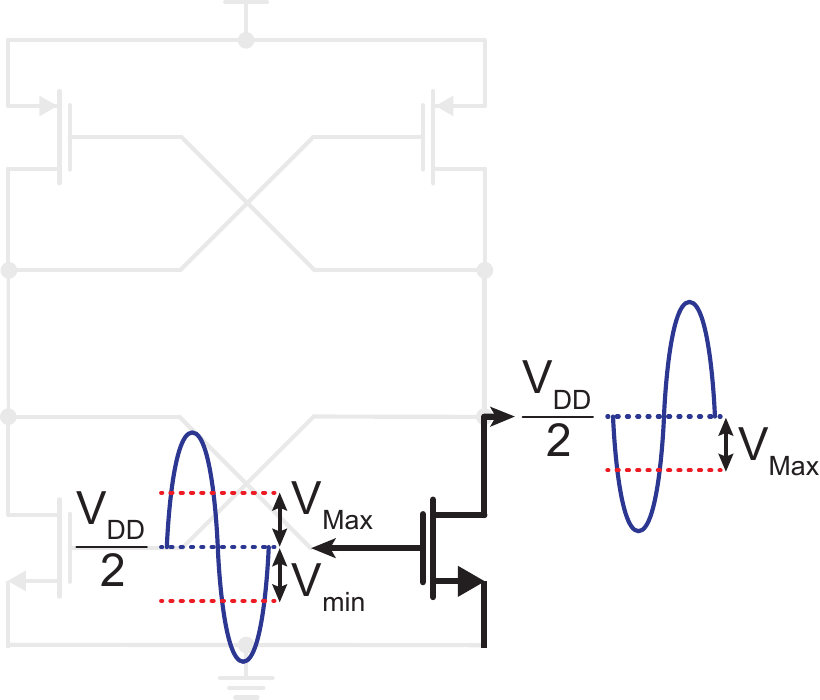}
    \label{fig:ngm_swing}}
    \subfloat[]{\includegraphics[width=1.7in]{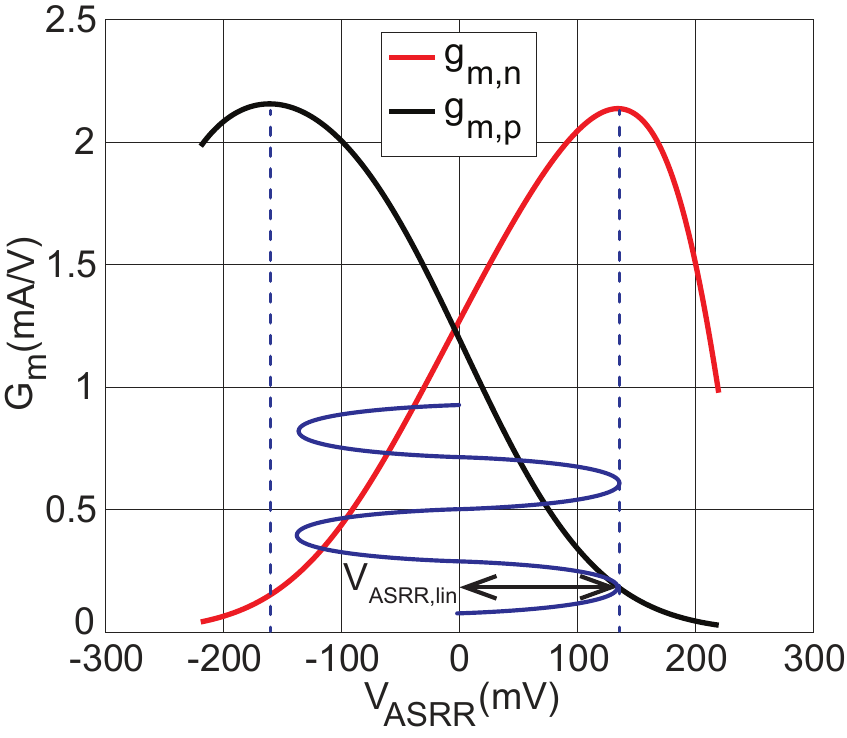}
    \label{fig:gmn_gmp}}    
    \subfloat[]{\includegraphics[width=1.7in]{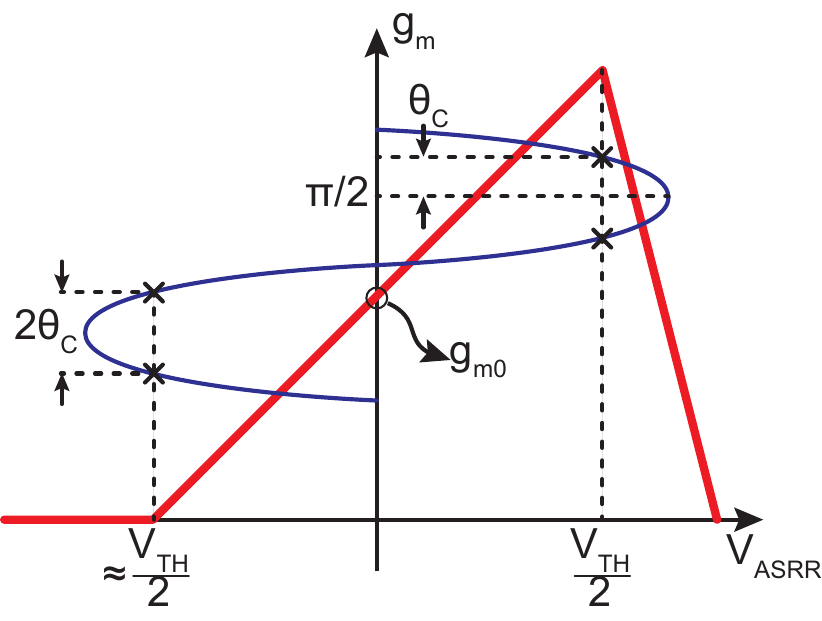}
    \label{fig:large_gm_sqr}}
    \subfloat[]{\includegraphics[width=1.7in]{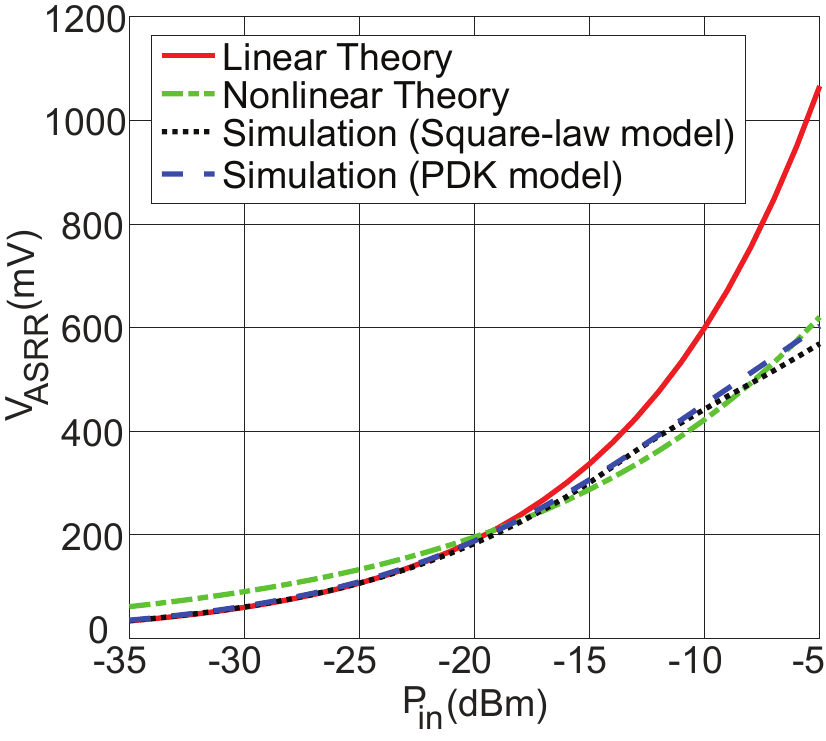}
    \label{fig:vasrr_pin}}
    
    \caption{(\subref{fig:ngm_swing}) Voltage swing limits across the ASRR. (\subref{fig:gmn_gmp}) Simulated $g_m$ of NMOS and PMOS, designed symmetrically to allow for maximum swing. (\subref{fig:large_gm_sqr}) The segmented curve used for calculating $g_{m,avg}$. (\subref{fig:vasrr_pin}) $V_{ASRR}$ across the $-g_m$ block, in an optimally coupled ASRR, built from square-law and PDK devices compared to values predicted by linear and nonlinear theories.}
    \label{fig:gm_comp}
\end{figure*}

\begin{figure}
    \centering
    \subfloat[]{\includegraphics[width=1.6in]{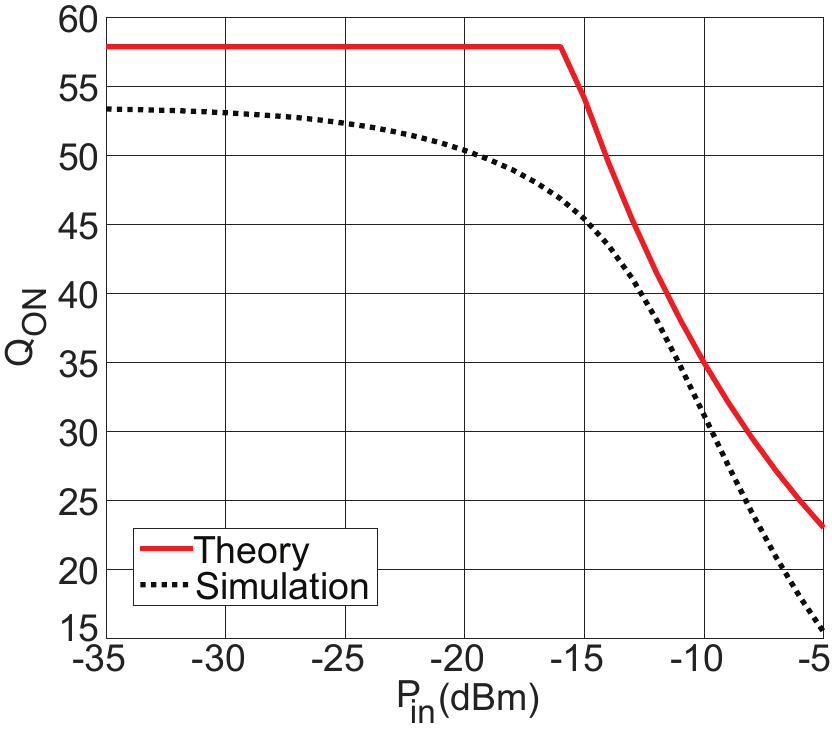}
    \label{fig:qout_pin}}
    \subfloat[]{\includegraphics[width=1.68in]{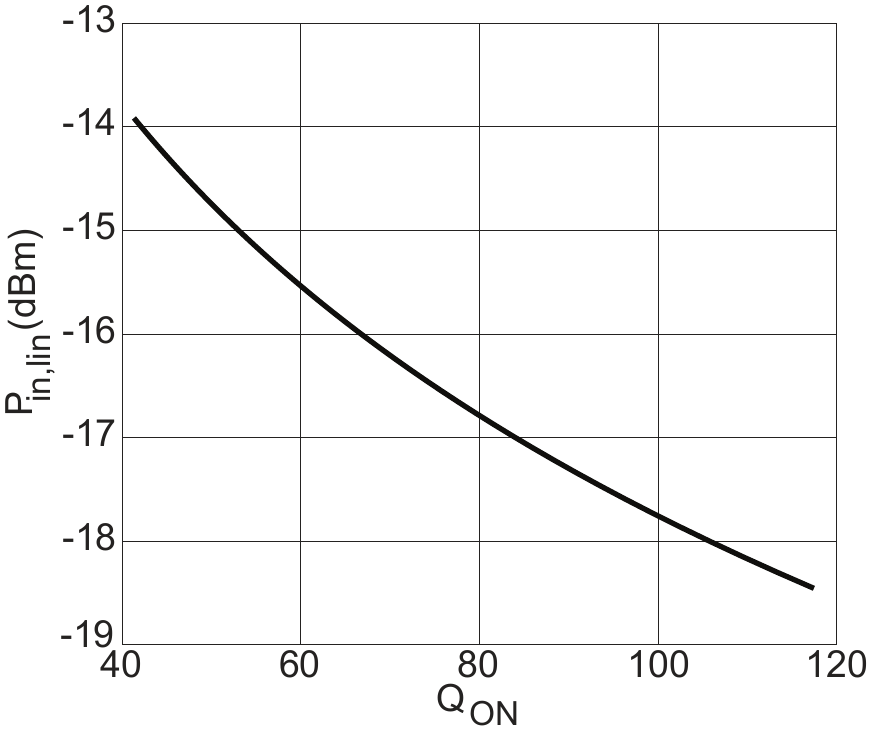}
    \label{fig:pinlin_qon}}
    \caption{(\subref{fig:qout_pin}) Nonlinear behavior of $Q_{ON}$ due to $-g_m$ compression. (\subref{fig:pinlin_qon}) $P_{in,lin}$ as a function of $Q_{ON}$.}
    \label{fig:q_nonlin_pinlin}
\end{figure}

The difference between these two frequencies is the usable frequency range of the ASRR, $\omega_{BW}=\omega_H-\omega_L\approx\frac{\omega_0}{Q_{ON}}$.
The higher the $S_{RES}$, the larger the detection gain, but the lower the maximum change in the resonance frequency that the pixel can experience before saturating. If the measurement frequency is set at the resonance frequency, then the maximum detection limit becomes $\omega_{BW}/2$.

\subsection{ASRR Nonlinearity}

In ASRRs, the negative resistance is a function of the voltage swing across the $-g_m$ block. This swing is ultimately determined by the resonator $Q_{ON}$ and the input power, $P_{in}$. By increasing either of these two parameters, the voltage swing across the resonator, $V_{ASRR}$, increases to a level where the small-signal circuit behaviors no longer hold, and devices start to operate in the cutoff and triode regions, thus diminishing the negative resistance effects. Assuming the internal nodes are biased at $\sim V_{DD}/2$, the onset of this phenomenon occurs roughly at $V_{TH}/2$ for the positive swing and $V_{DD}/2-V_{TH}$ for the negative swing, as shown in Fig. \ref{fig:gm_comp}(\subref{fig:ngm_swing}), outside of which the device enters the triode and cutoff regions, respectively. Assuming $V_{DD}/2-V_{TH}\approx V_{TH}/2$, the single-ended swing of the $-g_m$ is approximately $V_{TH}/2$ resulting in a differential linear swing of $V_{ASRR,lin}=V_{TH}$. This is confirmed via simulations for a symmetrically designed $-g_m$ block consisting of PDK devices, as shown in Fig. \ref{fig:gm_comp}(\subref{fig:gmn_gmp}), where the $V_{ASRR,lin}$ swing is limited to $\sim400mV$. The nonlinearity effects in the device transconductance can be modeled using square-law relationships as shown in Fig. \ref{fig:gm_comp}(\subref{fig:large_gm_sqr}). Here, only one device is considered due to symmetry, and the device $g_m$ is split into three different cutoff, saturation, and triode regions. $g_{m,avg}$ is defined as the average transconductance of the device over one signal cycle, i.e., $V_{ASRR}sin(\omega_{0} t)$. So long as the device remains in the saturation region, $g_{m,avg}=g_{m0}$, the DC transconductance of the device. However, if the $V_{ASRR}>V_{TH}/2$, as depicted in Fig. \ref{fig:gm_comp}(\subref{fig:large_gm_sqr}), $g_{m,avg}$ can be calculated by a three-segment integration, one over the saturation region, which amounts to $g_{m0}$, and two others over the triode and cut-off regions. These calculations are performed in Appendix \ref{appb}, where (\ref{eq_gm_avg}) and (\ref{eq_gm_avg_aprx}) give exact and approximate expressions for $g_{m,avg}$, respectively.

At resonance, this swing shows up across the ASRR parallel impedance, $R_{ASRR}$. The most significant effect of transconductance nonlinearity is the degradation in $Q_{ON}$, and, consequently, the pixel sensitivity. In what follows, the relationships between the input power to the host transmission line, $P_{in}$, $V_{ASRR}$, and the nonlinear behavior of the ASRR quality factor, denoted by $Q_{ON,nonlin}$, are studied.

Once again, the equivalent circuit in Fig. \ref{fig:srr_cpl}(\subref{fig:srr_cpl_tl_lc_2}) is employed, while noting that the powers consumed in the equivalent and actual SRRs are equal. The power consumed by the equivalent SRR can be calculated as

\begin{equation}
    P_{SRR'}=(1-S_{11}^2-S_{21}^2)P_{in}=\frac{4R_{SRR}'Z_0}{(R_{SRR}'+2Z_0)^2}P_{in}
    \label{eq_psrr'}
\end{equation}

However, this power is equal to $P_{SRR}$ due to circuit equivalency, therefore

\begin{equation}
    P_{SRR}=\frac{V_{SRR}^2}{2R_{SRR}}=\frac{4R_{SRR}'Z_0}{(R_{SRR}'+2Z_0)^2}P_{in}
    \label{eq_psrr}
\end{equation}

For an optimum coupled resonator, (\ref{eq_psrr}) simplifies to $P_{SRR}=4/9P_{in}$. Therefore, the voltage swing across an ASRR can be calculated as

\begin{equation}
    V_{ASRR,lin}=\sqrt{\frac{8}{9}R_{ASRR}P_{in}}=\sqrt{\frac{8}{9}\omega_{0}L_{SRR}Q_{ON}P_{in}}
    \label{eq_vasrr}
\end{equation}

For a given $Q_{ON}$, (\ref{eq_vasrr}) only holds when $P_{in}$ is below a level such that the $-g_m$ operates linearly. From the above discussion, this occurs when $V_{ASRR,lin}=V_{TH}$ and thus
\begin{equation}
    P_{in,lin}=\frac{9}{8}\frac{V_{TH}^2}{\omega_{0}L_{SRR}Q_{ON}}
    \label{eq_pin_lin}
\end{equation}

For $P_{in}>P_{in,lin}$, $Q_{ON}$ starts to enter a nonlinear domain and degrade due to $g_m$ compression and can be approximated from 
\begin{equation}
    Q_{ON,nonlin}=\frac{Q_{OFF}}{1-g_{m,avg}R_{SRR}}
    \label{eq_qon_nonlin}
\end{equation}
where $g_{m,avg}$ is derived in Appendix \ref{appb} (\ref{eq_gm_avg_aprx}). These derivations are verified using large-signal simulations, where $P_{in}$ is increased and $V_{ASRR}$ is measured and compared with the values computed from (\ref{eq_vasrr}). This comparison is illustrated by Fig. \ref{fig:gm_comp}(\subref{fig:vasrr_pin}), where the linear and nonlinear theories use $Q$ values calculated from (\ref{eq_qon}) and (\ref{eq_qon_nonlin}), respectively, and also $V_{ASRR}$ is measured for both $-g_m$ blocks made from both square-law and PDK devices. What is clear from these comparisons is that the linear theory fails to capture the transconductance compression as the input power increases, and the $V_{ASRR}$ for input powers above $P_{in,lin}$ starts to deviate significantly from the actual values obtained from simulations. These discrepancies, however, are resolved using the nonlinear theory. In Fig. \ref{fig:q_nonlin_pinlin}(\subref{fig:qout_pin}), the simulated $Q_{ON,nonlin}$ is compared with the values calculated from (\ref{eq_qon_nonlin}). Since a higher $P_{in}$ results in a higher conversion gain in the phase detector, setting $P_{in}=P_{in,lin}$ to the onset of the nonlinear region is the best compromise. The values of $P_{in,lin}$ as a function of $Q_{ON}$ are calculated from (\ref{eq_pin_lin}) and plotted in Fig. \ref{fig:q_nonlin_pinlin}(\subref{fig:pinlin_qon}), which guides the selection of input power once a $Q_{ON}$ value is chosen.

\begin{figure}
    \centering
    \includegraphics[width=3in]{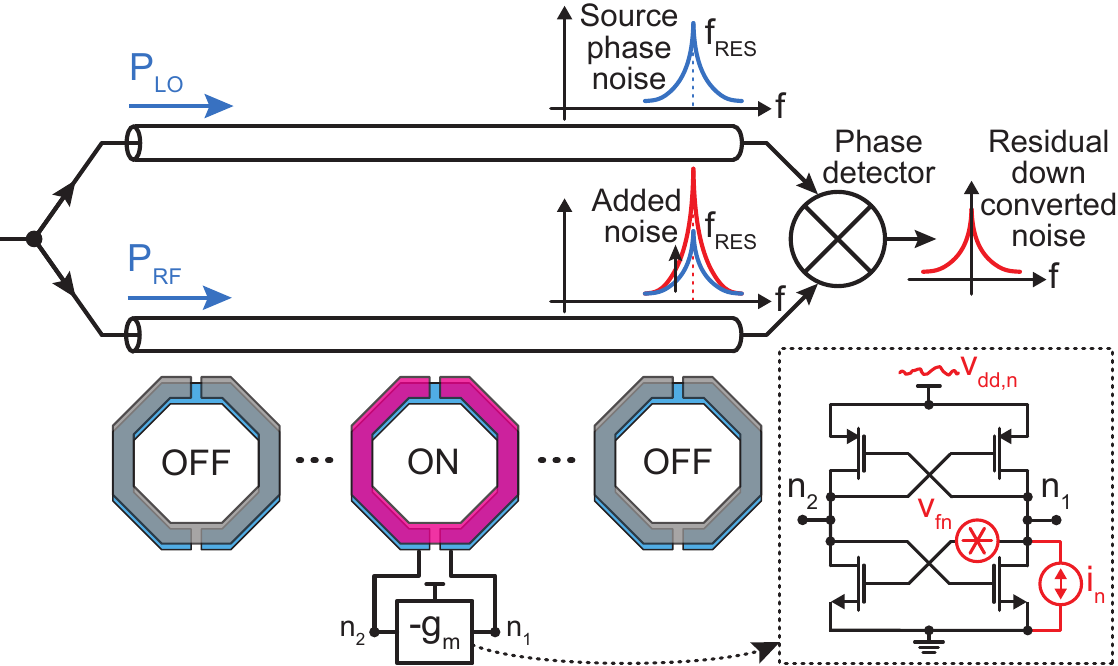}
    \caption{Various noise sources contributing to the output noise.}
    \label{fig:noise_overview}
\end{figure}

\begin{figure*}
    \centering
    \subfloat[]{\includegraphics[width=1.8in]{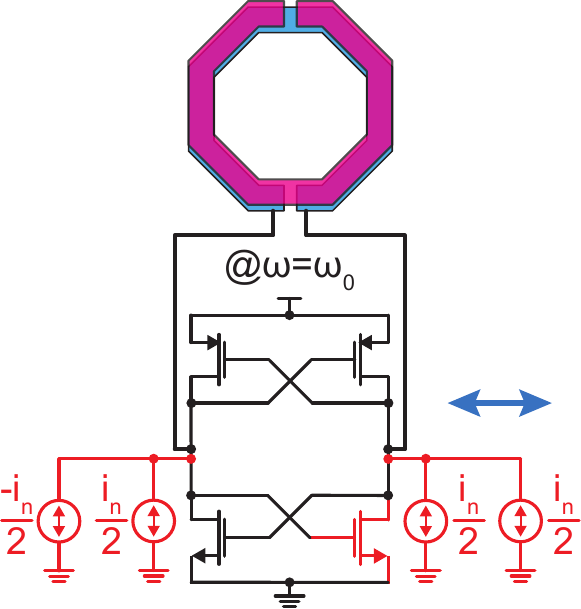}
    \label{fig:asrr_white_noise_1dev}}
    \subfloat[]{\includegraphics[width=0.725in]{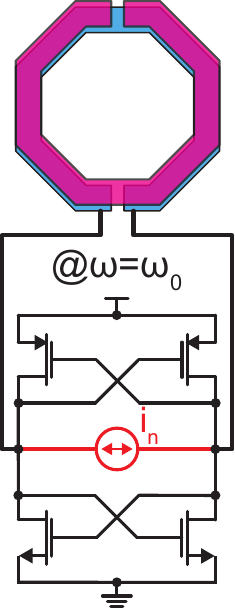}
    \label{fig:asrr_white_noise}}
    \subfloat[]{\includegraphics[width=1.65in]{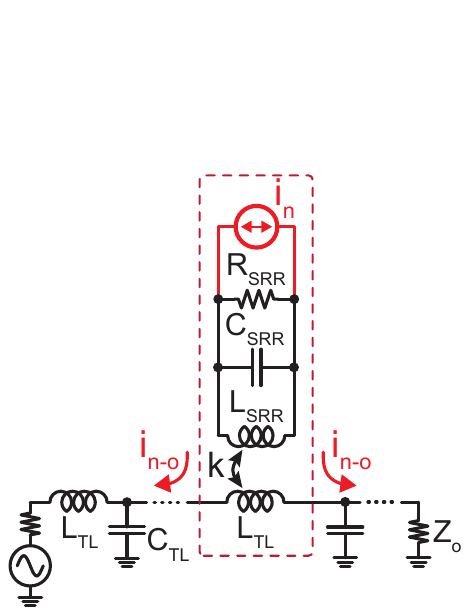}
    \label{fig:srr_lc_noise}}
    \subfloat[]{\includegraphics[width=1.65in]{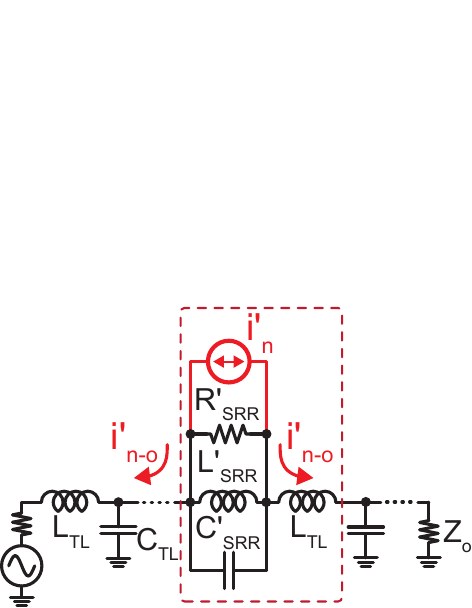}
    \label{fig:srr_lc_eq_noise}}
    \subfloat[]{\includegraphics[width=1.03in]{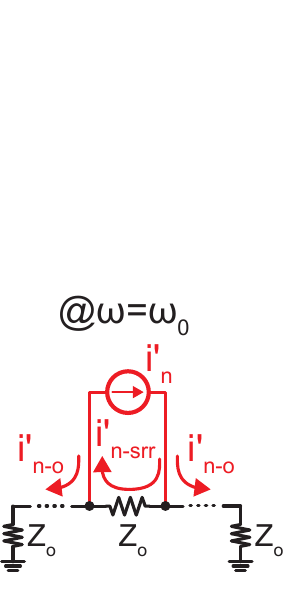}
    \label{fig:srr_eq_atres_noise}}
    
    \caption{(\subref{fig:asrr_white_noise_1dev}) The decomposition of white noise of a single device to differential and common-mode components. (\subref{fig:asrr_white_noise}) The total noise of all four devices that appears across ASRR. (\subref{fig:srr_lc_noise}) The total noise across the LC model. (\subref{fig:srr_lc_eq_noise}) The total noise across the equivalent LC model. (\subref{fig:srr_eq_atres_noise}) The noise flow at resonance.}
    \label{fig:asrr_noise_1}
\end{figure*}

\section{Noise in ASRR}\label{sec_noise_asrr}

Although an ASRR resembles a CMOS LC oscillator, it never oscillates, as its loop gain is intentionally kept below unity. As a result, the conventional noise theories developed for oscillators are not applicable here, and understanding the effect of various noise mechanisms requires its own treatment.

Fig. \ref{fig:noise_overview} illustrates the various noise sources that could affect the output signal. As will be shown in Section \ref{sec_inp_pnoise}, the phase detector cancels out the correlated phase noise between the $LO$ and $RF$ signals. As a result, the goal here is to analyze the effects of the pixel white, flicker, and supply noise on the output phase noise.

Here, a somewhat unconventional approach to low-frequency noise analysis is taken. First, the fluctuations of the \textit{impedance} of the $-g_m$ block due to each noise source are calculated. Next, the effect of impedance on the resonance characteristics is derived. Finally, the relationships between the resonance characteristics and the output phase noise are established, which collectively enable one to compute the output phase noise. It must be noted that the sensing bandwidth is relatively small ($\sim$1kHz), as measuring cells and tissues can be done slowly, and the phase detector also acts as a direct conversion receiver (zero IF). Therefore, the contribution and upconversion of the noise sources to $RF$ close-in phase noise is of primary interest.

For an optimally coupled SRR, the first quantity, which relates $S_{RES}$ to the ASRR impedance around resonance, $R_{ASRR}$, is calculable from (\ref{eq_dphi_dwdr_2}) as follows

\begin{equation}
    \frac{dS_{RES}}{dR_{ASRR}}=\frac{10}{9}C_{SRR}.
    \label{eq_dsres_dr_2}
\end{equation}

The second parameter captures the changes in the ASRR impedance as a function of changes in the transconductance and can be calculated from (\ref{eq_r_asrr}) as
\begin{equation}
    \frac{dR_{ASRR}}{dg_m}=\frac{R_{SRR}^2}{(1-g_m R_{SRR})^2}.
    \label{eq_drasrr_dgm}
\end{equation}

In the rest of this section, the contributions of each noise source to the total transconductance of the $-g_m$ block, and subsequently to the output phase noise, are derived.

\subsection{Device White Noise}\label{sec_wnoise}

\begin{figure}[t]
    \centering
    \subfloat[]{\includegraphics[width=1.68in]{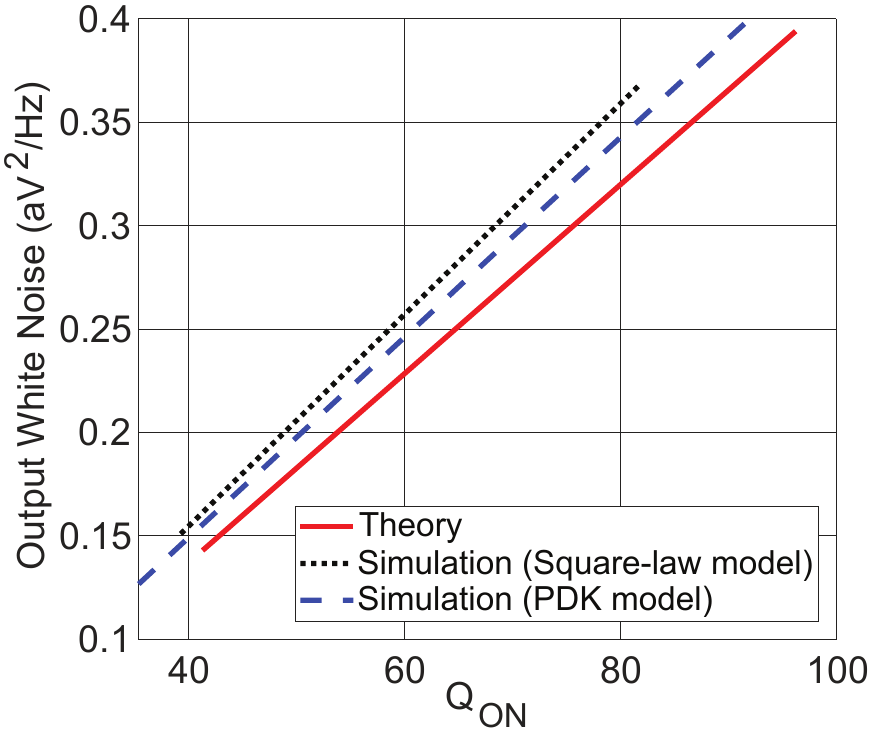}
    \label{fig:wnoise_qon_sim}}
    \subfloat[]{\includegraphics[width=1.68in]{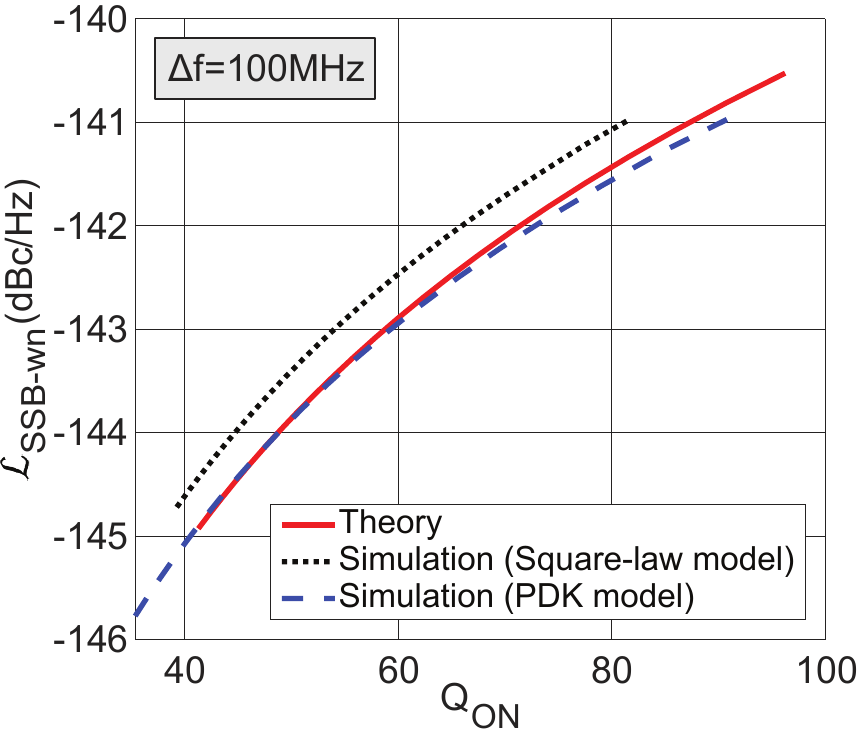}
    \label{fig:pnoise_wnoise_qon_sim}}

    \caption{(\subref{fig:wnoise_qon_sim}) Output noise at resonance frequency due to device white noise as a function of $Q_{ON}$. (\subref{fig:pnoise_wnoise_qon_sim}) Output phase noise at $\Delta f$=100MHz offset as a function of $Q_{ON}$.}
    \label{fig:asrr_wnoise_sim}
\end{figure}

\begin{figure*}
    \centering
    \subfloat[]{\includegraphics[width=2.1in]{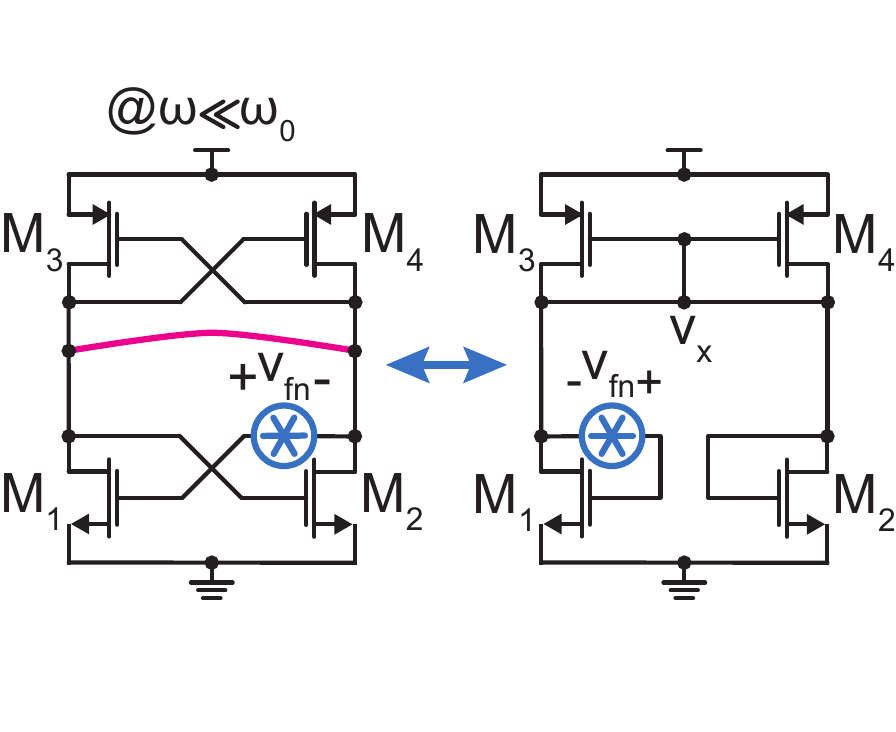}
    \label{fig:ngm_flicker_noise}}
    \subfloat[]{\includegraphics[width=1.7in]{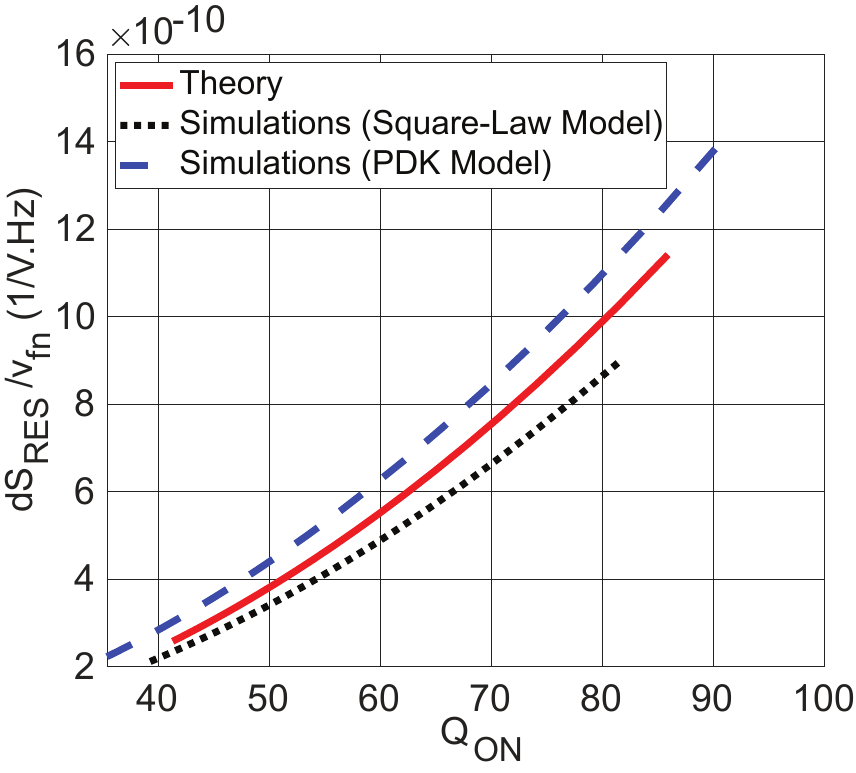}
    \label{fig:dsres_vfn_sim}}
    \subfloat[]{\includegraphics[width=1.4in]{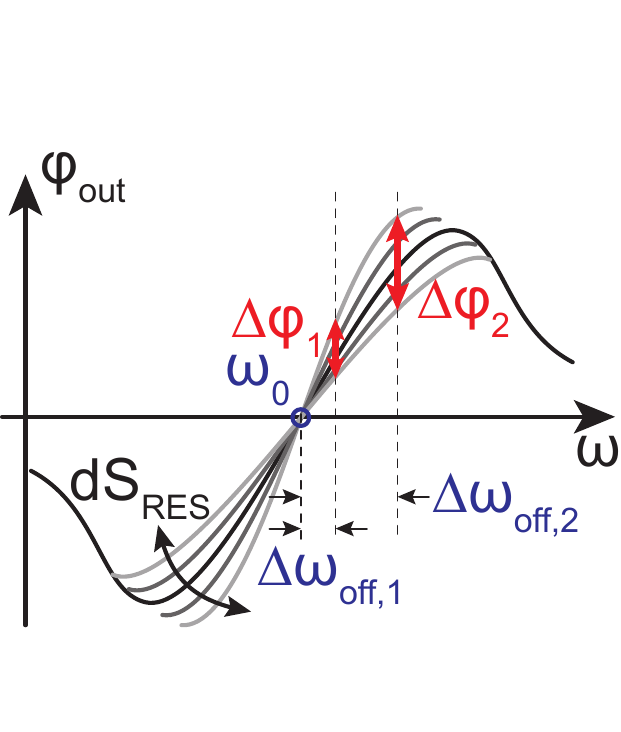}
    \label{fig:asrr_sres_noise}}
    \subfloat[]{\includegraphics[width=1.7in]{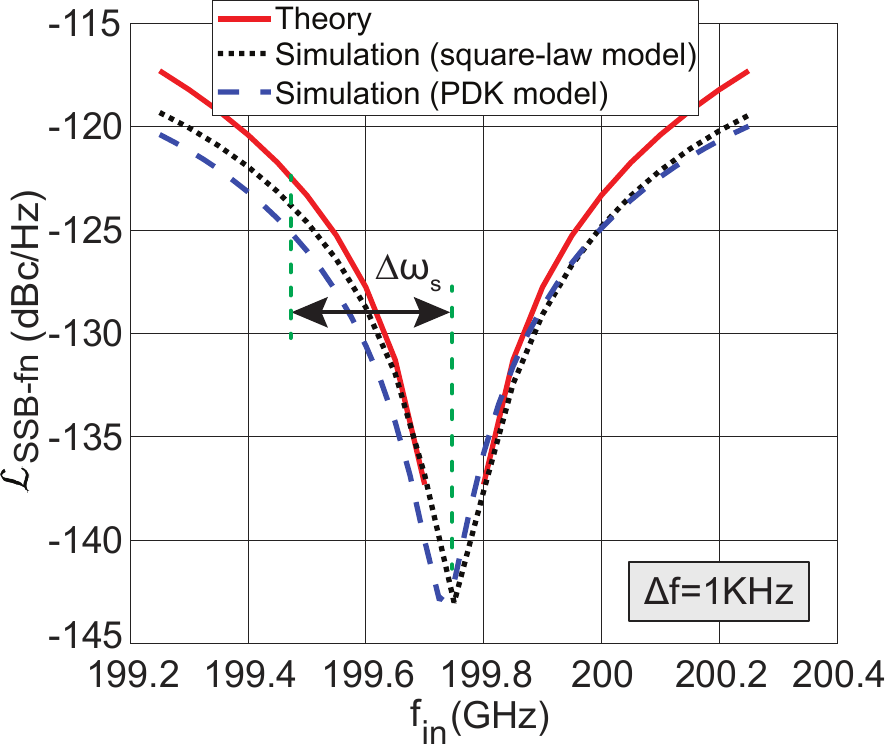}
    \label{fig:pnoise_flicker_fin}}    

    \caption{(\subref{fig:ngm_flicker_noise}) The circuit that interacts with flicker noise, $v_{fn}$. (\subref{fig:dsres_vfn_sim}) The transfer function through which $v_{fn}$ affects $S_{RES}$ for different $Q_{ON}$ values. (\subref{fig:asrr_sres_noise}) The effect of offset frequency on the amount of error that $dS_{RES}$ can cause in $\varphi_{out}$. (\subref{fig:pnoise_flicker_fin}) Theoretical and simulated output phase noise due to flicker noise at 1kHz offset.}
    \label{fig:asrr_noise_2}
\end{figure*}

The circuit that the device white noise, $i_n$, affects is drawn in Fig. \ref{fig:asrr_noise_1}(\subref{fig:asrr_white_noise_1dev}). Here, due to symmetry, the channel noise of only one device is considered, which can be extended to all four devices. Moreover, the noise is decomposed into common-mode and differential components and applied to both internal nodes, $n_{1,2}$. Only the differential component can flow through the resonator and introduce an error in the output signal. Alternatively, the common-mode component is rejected by its inability to induce a signal on the host transmission line and thereby perturb the output. As a result, only 1/2 of the instantaneous current noise or 1/4 of the noise power enters the calculations. Assuming the channel noise of four devices are equal, and extending this analysis to all four devices, one ends up with the equivalent circuit shown in \ref{fig:asrr_noise_1}(\subref{fig:asrr_white_noise}), where the total noise contribution of all devices shows up as a differential noise source, whose power is equal to that of a single device.

In the equivalent LC model, this noise source is added across the resonator as shown in Fig. \ref{fig:asrr_noise_1}(\subref{fig:srr_lc_noise}). In this model, part of the noise circulates inside the resonator, while the rest flows equally to the input and output ($i_{n-o}$). In the equivalent resonator model (SRR$'$), an equivalent noise current, $i_{n}^{'}$, can be placed across the resonator, which injects the currents $i_{n-o}^{'}$ to the input and output, as demonstrated in Fig. \ref{fig:asrr_noise_1}(\subref{fig:srr_lc_eq_noise}). The relationship between $i_n$ and $i_{n}^{'}$ can be found based on the following argument: these two circuits are equivalent, which means that the noise powers, due to $i_n$ and $i_n^{'}$, inside the two resonators, as well as the source and termination impedances, must be equal. This means two things: 1- $i_{n-o}=i_{n-o}^{'}$, since the source and termination impedances are equal to $Z_0$ in both cases, and 2- at resonance, if $ai_n$ and $ai_n^{'}$ denote the fraction of the noise currents that flow through the resonators losses, i.e. $R_{SRR}$ and $R_{SRR}^{'}$, the noise powers inside the two resonators are equal and, for a 1 Hz bandwidth, given by

\begin{equation}
    R_{ASRR}\cdot a^2\overline{i_n^2}=R_{ASRR}'\cdot a^2\overline{i_n'^2}.
    \label{eq_noise_equiv}
\end{equation}

For an optimally coupled resonator $R_{ASRR}'=Z_0$, and thus the noise current density in the SRR$'$ can be expressed as 
\begin{equation}
    \overline{i_n'^2}=\frac{R_{ASRR}}{Z_0}\overline{i_n^2}.
    \label{eq_noise_srrp}
\end{equation}

Additionally, according to Fig. \ref{fig:asrr_noise_1}(\subref{fig:srr_eq_atres_noise}), at resonance, 1/3 of the instantaneous noise current circulates in the resonator, i.e. $a=1/3$, and the rest is equally split between input and output. Therefore, the output noise current density $\overline{i_{n-o}'^{2}}=1/9\overline{i_n'^{2}}$. Therefore, based on (\ref{eq_noise_srrp}), the total output voltage noise density due to the noise source $i_n$ is given by
\begin{equation}
    \overline{v_{n,out}^2}=\frac{1}{9}R_{ASRR}Z_0\overline{i_n^2}.
    \label{eq_out_noise_white1}
\end{equation}

Using the MOS channel noise equation $\overline{i_n^2}=4kT\gamma g_m$, (\ref{eq_out_noise_white1}) can be expressed as
\begin{equation}
    \overline{v_{n,out}^2}=\frac{1}{9}Q_{ON}\omega_{0}L_{SRR}Z_0(4kT\gamma g_m).
    \label{eq_out_noise_white2}
\end{equation}

Fig. \ref{fig:asrr_wnoise_sim}(\subref{fig:wnoise_qon_sim}) shows the simulation results of output noise density for an optimally coupled ASRR, for $-g_m$ built from square-law and PDK devices, compared to theoretical values derived from (\ref{eq_out_noise_white2}), where the $\gamma$ value was extracted from simulating a single device. Since the ASRR is operating in the linear regime and the input signal does not modulate the white noise originating from the ASRR, the noise from (\ref{eq_out_noise_white2}) is additive and thus splits evenly between AM and PM \cite{Darabi_rfic}. Therefore, the SSB phase noise can be calculated by taking the ratio between the PM sideband power and the carrier power as follows

\begin{equation}
    \mathlarger{\mathcal{L}}_{SSB,wn}(f)=10log\left(\frac{\frac{1}{2}\overline{v_{n,out^2}}}{50\Omega\cdot P_{DET}}\right).
    \label{eq_pnoise_wnoise}
\end{equation}
where $P_{DET}=P_{in}$-3.52dB for an optimally coupled ASRR. The simulation results, presented in Fig. \ref{fig:asrr_wnoise_sim}(\subref{fig:pnoise_wnoise_qon_sim}), show a close agreement between the phase noise values given by (\ref{eq_pnoise_wnoise}) and the simulated SSB phase noise at 100MHz offset frequency. A large offset frequency was chosen to ensure the results are not affected by the device flicker noise.

\subsection{Device Flicker Noise}

The low-frequency noise of the devices affects the resonator response differently. Both $\omega_{0}$ and $Q_{ON}$ can be modulated by changing the DC operating points of the negative resistance due to $1/f$ noise. The former is caused by variations in the device parasitic capacitances, which are quite negligible and will not be considered in our discussions. The latter, however, can be more significant and warrants further study.

At low frequencies, the resonator acts as a short, transforming the $-g_m$ block to four diode-connected devices. Fig. \ref{fig:asrr_noise_2}(\subref{fig:ngm_flicker_noise}) captures this, where again, due to the symmetry in this block, only the flicker noise of one device, $v_{fn}$, is considered, resulting in the two identical circuits, shown side-by-side. The small-signal internal voltage, $v_{x}$, induced by the noise can be found from the nodal equation

\begin{equation}
    g_{m1}(v_x+v_{fn})+(g_{m2}+g_{m3}+g_{m4})v_x=0
\end{equation}
which, assuming equal transconductance for all devices, yields $v_x=-\frac{v_{fn}}{4}$. Therefore, the small-signal gate-source voltage due to noise for each device is
\begin{subequations}
\begin{align}
    v_{gs1}&=\frac{3}{4}v_{fn}\\
    v_{gs2,3,4}&=-\frac{1}{4}v_{fn}
\end{align}
\label{eq_vgs_1234}    
\end{subequations}
which result in a change in device transconductance, denoted by $dg_{mi}\,(i=1,2,3,4)$. Furthermore, for two transconductances in series after undergoing two small arbitrary changes, $\delta_1$ and $\delta_2$, the total transconductance can be estimated from

\begin{equation}
    (g_{m}+\delta_1)||(g_{m}+\delta_2)\approx\frac{g_m}{2}+\frac{\delta_1+\delta_2}{4}.
    \label{dgm_parallel}
\end{equation}

To find the change in the resonator behavior as a result of $v_x$, the variations in the total transconductance of the block in terms of variations in the PMOS and NMOS transconductance are calculated using (\ref{dgm_parallel}) as

\begin{align}
    dg_m&\approx\left(\frac{dg_{m1}+dg_{m2}}{4}+\frac{dg_{m3}+dg_{m4}}{4}\right)
    \label{eq_dgm}
\end{align}
where
\begin{equation}
    \frac{dg_{mi}}{v_{gsi}}=K_i\left(\frac{W}{L}\right)_i(1+\lambda(2V_{GSi}-V_{TH}))
    \label{eq_dgmi}
\end{equation}
for a square law device. Utilizing (\ref{eq_vgs_1234}), (\ref{eq_dgm}) can be written as
\begin{align}
    dg_m&=\frac{1}{4}\left(\frac{dg_{mn}}{v_{gsn}}(\frac{3}{4}-\frac{1}{4})v_{fn}+\frac{dg_{mp}}{v_{gsp}}(\frac{1}{4}+\frac{1}{4})v_{fn}\right)\nonumber\\
    &=\frac{1}{8}\left(\frac{dg_{mn}}{v_{gsn}}+\frac{dg_{mp}}{v_{gsp}}\right)v_{fn}.
    \label{eq_dgm_vfn}
\end{align}
    
Using (\ref{eq_dsres_dr_2}), (\ref{eq_drasrr_dgm}), and (\ref{eq_dgm_vfn}), the variations in $S_{RES}$ due to the noise source $v_{fn}$ can be computed as follows

\begin{align}
    &dS_{RES}=\frac{dS_{RES}}{dR_{ASRR}}\cdot\frac{dR_{ASRR}}{dg_m}\cdot dg_m\nonumber\\
    &=\frac{5}{36}L_{SRR}Q_{ON}^2\left(\frac{dg_{mn}}{v_{gsn}}+\frac{dg_{mp}}{v_{gsp}}\right)v_{fn}.
    \label{eq_dsres_vfn}
\end{align}

The $1/f$ noise sensitivity parameter, $\alpha_{1/f}\triangleq5/36({dg_{mn}}/{v_{gsn}}+{dg_{mp}}/{v_{gsp}})$, is defined, which will be used in the following sections. Neglecting channel length modulation effects, $\alpha_{1/f}\approx 5/36(K_n(W/L)_n+K_p(W/L)_p)$.

Equation (\ref{eq_dsres_vfn}) was verified by simulations while sweeping $Q_{ON}$, applying a small offset $v_{fn}$, and measuring the ratio, $dS_{RES}/v_{fn}$. The results are shown in Fig. \ref{fig:asrr_noise_2}(\subref{fig:dsres_vfn_sim}), where the theoretical values from (\ref{eq_dsres_vfn}) are compared to simulation results for square-law and CMOS 28nm PDK devices. The output noise exhibits a quadratic dependence on $Q_{ON}$, posing a challenge if large boosted quality factors are targeted.

By referring to Fig. \ref{fig:asrr_noise_2}(\subref{fig:asrr_sres_noise}), one can see that $S_{RES}$ fluctuations, $dS_{RES}$, causes the output phase to pivot around $\omega_{0}$. This shows that at the resonance frequency, the output phase is immune to $dS_{RES}$, i.e. the phase error $\Delta\varphi=0$, and additionally, using the definition of $S_{RES}$ the phase error at $\Delta\omega_{s}$ away from the resonance, can be approximated as $\Delta\varphi=S_{RES}\Delta\omega_{s}$. From this relationship and (\ref{eq_dsres_vfn}), the output phase noise due to all four devices is given by

\begin{equation}
    \mathlarger{S}_{\varphi,fn}(\Delta f)=4\left(\frac{dS_{RES}}{v_{fn}}\right)^2\cdot\overline{v_{fn}^2}(\Delta f)\cdot\Delta\omega_{s}^2
    \label{eq_pnoise_fnoise}
\end{equation}
where $\overline{v_{fn}^2}(f)=K_f/f$, $K_f$ is the device flicker noise coefficient, $\Delta f$ is the offset frequency where the $1/f$ noise is measured. The larger the $\Delta\omega_{s}$, the greater the phase noise, as the phase response deviates from its nominal position by a larger amount. 

It is noteworthy that the white noise analysis in Section \ref{sec_wnoise} was performed in the voltage domain, and the calculated white noise PM sideband power should have been normalized to the carrier power. The flicker noise analysis provided by (\ref{eq_pnoise_fnoise}), however, is directly done in the phase domain, and the SSB phase noise is readily available from  $\mathcal{L}_{SSB,fn}=10log(S_{\varphi,fn})$. At a $\Delta f$=1kHz flicker noise offset, the phase noise is plotted in Fig. \ref{fig:asrr_noise_2}(\subref{fig:pnoise_flicker_fin}), where $K_f$ for square-law devices was extracted from PDK device simulations. These results depict the expected notch at the resonance frequency ($\Delta \omega_s=0$), where the phase noise decreases to the level of white noise, and also validate the theory at non-zero $\Delta \omega_s$ values.

\subsection{Supply Noise}

The $-g_m$ supply low-frequency noise also gets upconverted to the resonance frequency by modulating the transconductance of the block. A small-signal $v_{dd}$ on the supply voltage induces a $v_{dd}/2$ voltage shift in the internal nodes, $n_{1,2}$, which results in a change in $-g_m$ value and therefore $S_{RES}$.

The sensitivity of the transconductance of the device to the supply voltage can be approximated as

\begin{equation}
    \frac{dg_{mi}}{v_{dd}}=K_{i}(\frac{W}{L})_{i}(\frac{1}{2}+\frac{\lambda}{2}(V_{DD}-V_{TH})).
    \label{eq_dgm_dvdd}
\end{equation}

Given that the total transconductance of the block is given by $g_m=(g_{m,n}+g_{m,p})/2$, the variations in $S_{RES}$ due to  $v_{dd}$ can be computed using (\ref{eq_dsres_dr_2}), (\ref{eq_drasrr_dgm}), and (\ref{eq_dgm_dvdd}) as

\begin{align}
    \frac{dS_{RES}}{v_{dd}}=\frac{5}{9}L_{SRR}Q_{ON}^2\left(\frac{dg_{mn}}{v_{dd}}+\frac{dg_{mp}}{v_{dd}}\right).
    \label{eq_dsres_dvdd}
\end{align}

If the supply is provided by an off-chip low-noise LDO, such as ADI's \textit{LT3045-1}, simulations show that the phase noise caused by the supply is negligible compared to device noise contribution and thus does not merit further attention. 

\subsection{Input Signal Phase Noise}\label{sec_inp_pnoise}

\begin{figure}[t]
    \centering
    \subfloat[]{\includegraphics[width=1.7in]{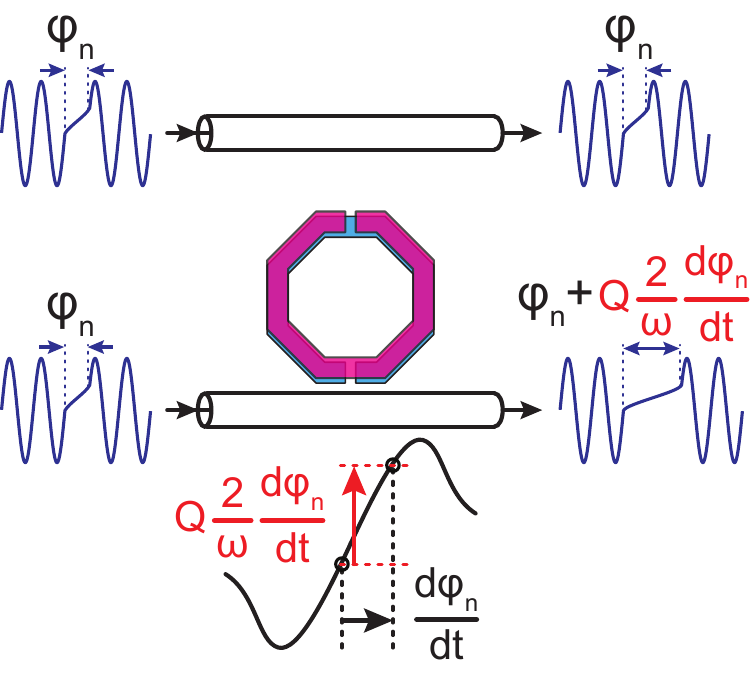}
    \label{fig:input_pnoise_sch}}    
    \subfloat[]{\includegraphics[width=1.64in]{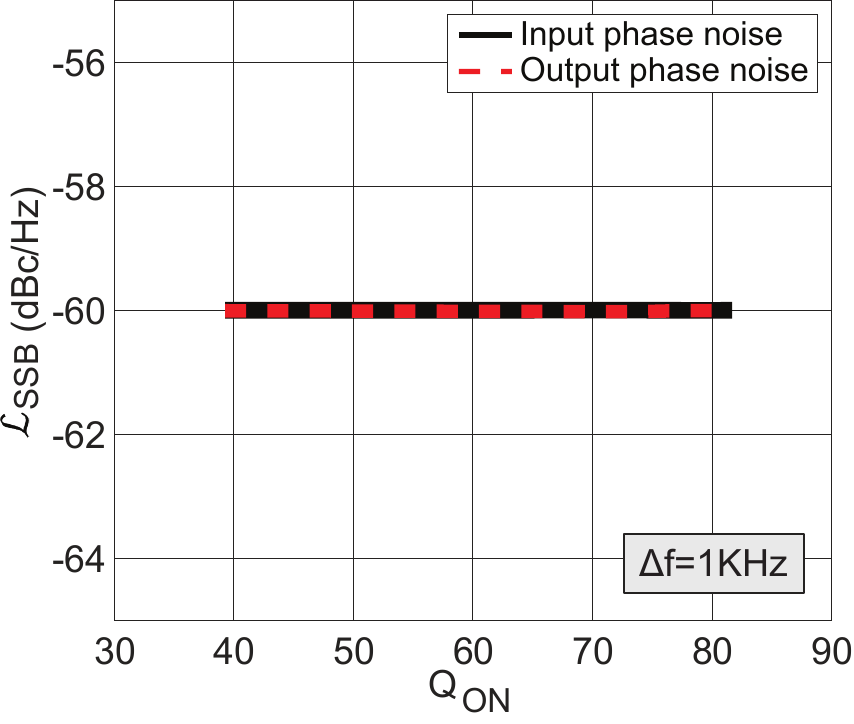}
    \label{fig:input_pnoise_sim}}
    \caption{(\subref{fig:input_pnoise_sch}) Phase noise propagation through a transmission line and an SRR-loaded transmission line. (\subref{fig:input_pnoise_sim}) Simulations of the input and output phase noise of a loaded transmission line.}
    \label{fig:input_pnoise_1}
\end{figure}

The noise on a signal propagating through a transmission line, in the absence of the resonator, directly appears at the output, albeit with some delay. However, as illustrated in Fig. \ref{fig:input_pnoise_1}(\subref{fig:input_pnoise_sch}), in the presence of a resonator, the output phase accumulates an extra phase, which depends on the rate at which the input phase changes. In other words, a sudden shift, $\varphi_n$, due to the input phase noise not only shows up in the output phase, but also gives rise to an extra phase shift through the resonator, resulting in a total output phase error of

\begin{equation}
    \varphi_{out}=\varphi_{n}+Q_{ON}\frac{2}{\omega_0}\frac{d\varphi_{n}}{dt}
    \label{eq_phi_out}
\end{equation}

Expressing (\ref{eq_phi_out}) in the s-domain in terms of the input and output phase noise results in

\begin{equation}
    S_{\varphi_{out}}=S_{\varphi_{in}}\left|1+sQ_{ON}\frac{2}{\omega_0}\right|^2
    \label{eq_inout_phasenoise}
\end{equation}

Equation (\ref{eq_inout_phasenoise}) implies that around the resonance, the output phase noise is identical to the input phase noise and also that there is a zero in the transfer function at the resonator bandwidth, $2Q_{ON}/\omega_0$. The output phase noise is, however, unaffected by this zero, as the pixel is always operated within the resonator bandwidth. Simulation results in Fig. \ref{fig:input_pnoise_1}(\subref{fig:input_pnoise_sim}) confirm that the input and output phase noise for different $Q_{ON}$ at low offsets are identical. This also confirms our original claim that phase detection suppresses the phase noise from the sub-THz source.

\begin{figure*}[t]
    \centering
    \subfloat[]{\includegraphics[width=1.8in]{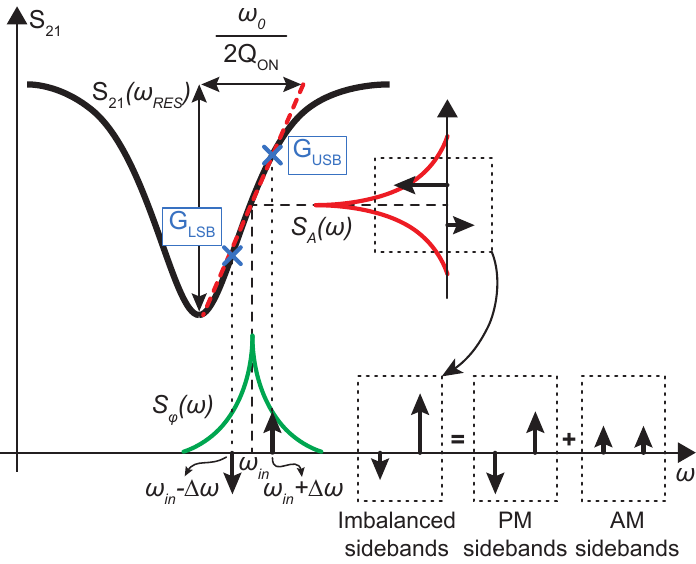}
    \label{fig:pm_to_am}}
    \subfloat[]{\includegraphics[width=1.7in]{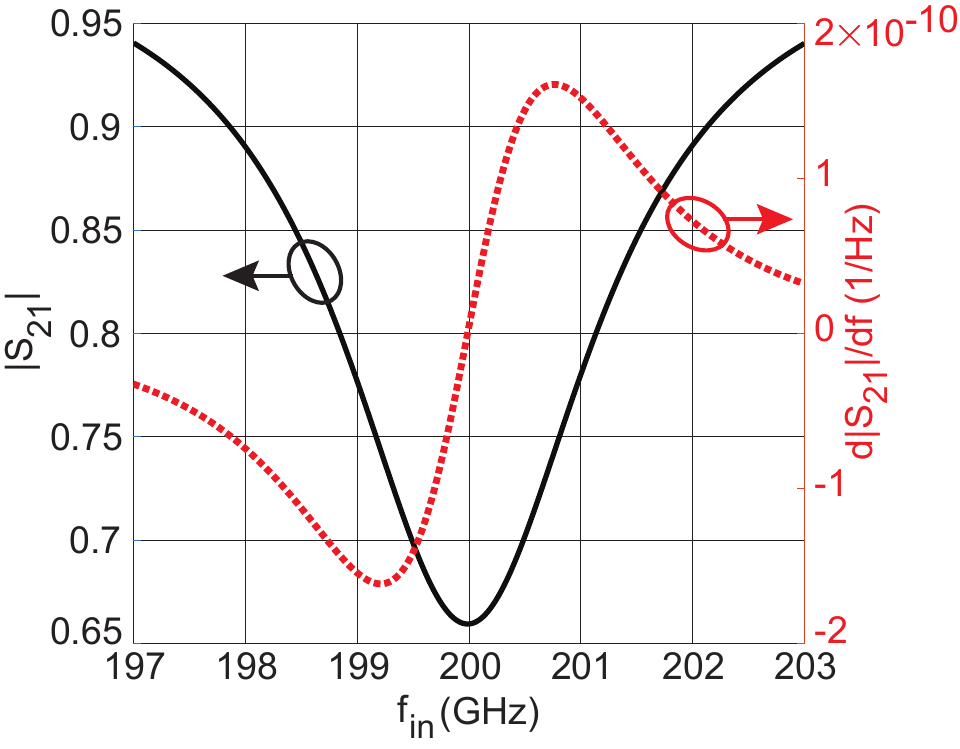}
    \label{fig:s21_ds21_df}}    
    \subfloat[]{\includegraphics[width=1.7in]{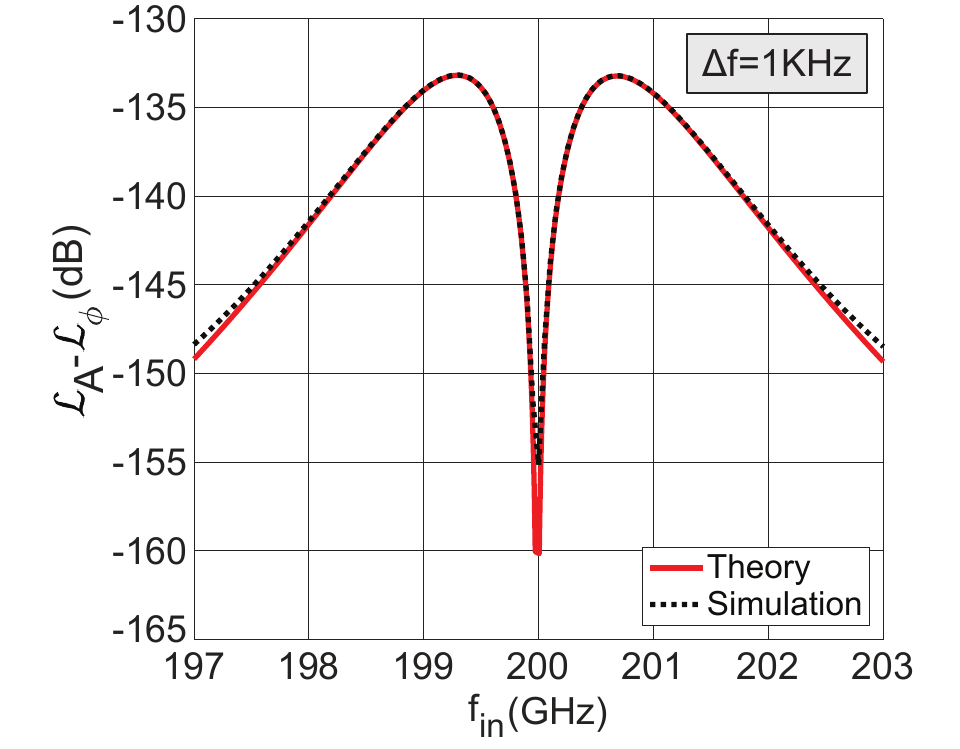}
    \label{fig:pm_to_am_fin}}
    \caption{(\subref{fig:pm_to_am}) PM sidebands give rise to AM sidebands due to the imbalances USB and LSB experience. (\subref{fig:s21_ds21_df}) $|S_{21}|$ and $d|S_{21}|/df$ of an optimally coupled ASRR. (\subref{fig:pm_to_am_fin}) Simulation vs. theoretical results of PM-to-AM conversion gain.}
    \label{fig:input_pnoise_2}
\end{figure*}

A signal traveling through a resonator experiences PM-to-AM conversion, which can also be derived analytically. This conversion peaks at the inflection points of the magnitude of $S_{21}$ and goes to zero at the resonance frequency. As pointed out in \cite{wang_gasspec_tbiocas2018}, the two PM sidebands experience an imbalance after interacting with the resonator, as shown in Fig. \ref{fig:input_pnoise_2}(\subref{fig:pm_to_am}). The resulting sidebands can be decomposed into two pairs of PM and AM sidebands, where the PM sidebands are roughly identical to those of the input, as discussed above, and the AM sidebands are thus given by half the difference between the output sidebands. Assuming the input signal contains purely PM sidebands, utilizing narrow-band approximation, they can be expressed as

\begin{align}
     V_{in}(t)&=Acos(\omega_{in}t)-\frac{\Delta\varphi A}{2}cos((\omega_{in}-\Delta\omega)t-\theta)\nonumber\\
    &+\frac{\Delta\varphi A}{2}cos((\omega_{in}+\Delta\omega)t+\theta)
    \label{eq_fm_sb}
\end{align}
where the second and third terms on the right-hand side are lower and upper sidebands (LSB and USB), respectively, and $\theta$ is an arbitrary phase difference between the fundamental tone and the sidebands. The average of the phases of the two sidebands is $\pi/2$, making their summation perpendicular to the fundamental tone, thus only modulating the phase \cite{Darabi_rfic}. Additionally, $\Delta\varphi A/2$ captures the phase noise at a frequency offset $\Delta\omega$, whose power normalized to the fundamental tone power is the input phase noise, $\mathcal{L}_{\varphi}(\Delta\omega)$. 

After going through the ASRR, the two sidebands experience slightly different gains, denoted by $G_{USB}$ and $G_{LSB}$, which can be approximated by the Taylor expansion of $S_{21}$ around $\omega_{in}$ as

\begin{align}
    G_{USB}(\omega_{in}+\Delta\omega)&\approx S_{21}(\omega_{in})+\frac{d|S_{21}|}{d\omega}\cdot\Delta\omega\label{eq_usb}\\
    G_{LSB}(\omega_{in}-\Delta\omega)&\approx S_{21}(\omega_{in})-\frac{d|S_{21}|}{d\omega}\cdot\Delta\omega\label{eq_lsb}
\end{align}

The AM sidebands can be calculated from (\ref{eq_fm_sb}), (\ref{eq_usb}), and (\ref{eq_lsb}) by eliminating the PM sidebands as

\begin{align}
    &V_{AM}(\omega_m)=\frac{1}{2}\left(V_{USB}(\omega_{in}+\Delta\omega)+V_{LSB}(\omega_{in}-\Delta\omega)\right)\nonumber\\
    &=\frac{1}{2}\left(G_{USB}\frac{\Delta\varphi A}{2}+G_{LSB}\frac{-\Delta\varphi A}{2}\right)=\frac{d|S_{21}|}{d\omega}\Delta\omega\frac{\Delta\varphi A}{2}.
\end{align}

From here, the output AM noise can be computed by taking the ratio between the AM sideband power and the carrier power, which results in

\begin{equation}
    \mathcal{L}_{A}(\Delta\omega)=20log\left(\frac{\frac{d|S_{21}|}{d\omega}\Delta\omega}{S_{21}(\omega_{in})}\right)+\mathcal{L}_{\varphi}(\Delta\omega).
    \label{eq_am_noise}
\end{equation}

The exact calculation of the $d|S_{21}|/d\omega$ is intractable. However, one can resort to simulations to find its maximum value, which occurs at the inflection points of the $S_{21}$ curve, as depicted in Fig. \ref{fig:input_pnoise_2}(\subref{fig:s21_ds21_df}). Fig. \ref{fig:input_pnoise_2}(\subref{fig:pm_to_am_fin}) shows the simulated PM-to-AM conversion gain as well as the calculated gain from (\ref{eq_am_noise}) using simulated values of $d|S_{21}|/df$, showing a great agreement between the theory and simulations. Although the architecture in this work is relatively immune to AM noise, these calculations become more pronounced when amplitude detection is performed.

\section{SNR Calculations and Design Guidelines}\label{sec_snr}

Simulating the output phase noise reveals that at a 1kHz offset, the primary noise contributors are the NMOS and PMOS flicker noise in the $-g_m$ block, followed by the supply noise and device white noise. Therefore, in the following discussion, only the flicker noise is considered. From (\ref{eq_pnoise_fnoise}), the RMS noise up to a 1kHz bandwidth, is given by
\begin{align}
    N_{fn}=v_{fn,rms}\left(\frac{dS_{RES}}{v_{fn}}\right)\Delta\omega_{s}
\end{align}
where $v_{fn,rms}$ is the device RMS flicker noise, and here $\Delta\omega_{s}$ is the shift in the resonance frequency due to the sample. For a change in the resonance frequency due to $\Delta C$, the output SNR can be evaluated from

\begin{align}
    SNR_{\Delta C}&=\frac{S_{RES}\Delta\omega_{s}}{v_{fn,rms}\left(\frac{dS_{RES}}{v_{fn}}\right)\Delta\omega_{s}}=\frac{1}{6\alpha_{1/f}v_{fn,rms}R_{ASRR}}.
    \label{eq_snr_re}
\end{align}
Likewise, the SNR due to $\Delta R$ can be written as
\begin{equation}
    SNR_{\Delta R}=\frac{\Delta S_{RES}\Delta\omega_{s}}{v_{fn,rms}\left(\frac{dS_{RES}}{v_{fn}}\right)\Delta\omega_{s}}=\frac{5\Delta R}{18\alpha_{1/f}v_{fn,rms}R_{SRR}^2}.
    \label{eq_snr_im}
\end{equation}

One key observation from (\ref{eq_snr_re}) and (\ref{eq_snr_im}) is that the SNR in both cases is independent of the shift in the resonance frequency due to $\Delta C$\footnote{Here, the SNR calculations only take the 1/f noise into account. For a small $\Delta C$, the SNR becomes limited by the white noise, which requires a separate treatment.}. Based on these findings, the following design guidelines for maximizing the SNR under minimum power consumption must be followed:
\\1- To maximize $SNR_{\Delta C}$, $R_{ASRR}$ must be minimized. The minimum value is determined by $Q_{ON,min}$, which is dictated by the coupling conditions provided by (\ref{eq_qon_min}).
\\2- To maximize $SNR_{\Delta R}$, $R_{SRR}$ must be minimized. However, the smaller $R_{SRR}$, the larger the required $g_m$ to boost the quality factor to $Q_{ON,min}$, which increases the power consumption. Therefore, for minimum power, $R_{SRR}$ must be lowered until $SNR_{\Delta R}$ is met.
\\3- Accordingly, $L_{SRR}$ must be designed such that it meets the imaging resolution and also provides the requisite $R_{SRR}$.
\\4- By choosing $L_{SRR}$ and $R_{SRR}$, the $g_m$ needed to achieve $Q_{ON,min}$ can be found from (\ref{eq_qon}).
\\5- The $g_m$ is implemented symmetrically, and is thus self-biased through the resonator at $V_{DD-g_m}/2$. The aspect ratio of the devices, $W/L$, can be found from the $-g_m$ value.
\\6- Assuming the device capacitive loading dominates the parasitic capacitance of the resonator, and from the resonance frequency, an equation for $W\times L$ can be derived.
\\7- From the two equations found above, the device $W$ and $L$ can be calculated.
\\8- At this point, $\alpha_{1/f}$ is determined based on the device parameters. Thus, if needed, the SNR can only be further improved by scaling both $W$ and $L$ by the same factor, which decreases $v_{fn,rms}$, at the expense of introducing more parasitic capacitance, which must be compensated for by reducing $L_{SRR}$ to restore the resonance frequency.

\section{Conclusion}\label{sec_conclusion}
In this paper, the analysis and design of ASRRs were presented. Two key features, namely the tunable boosted quality factor and the switchability, enable the dense integration of ASRRs in scalable 2D arrays. The near fields generated by these arrays are employed for high-resolution imaging of biological tissues, which find applications in disease diagnostics. The resonators are highly sensitive to both the real and imaginary parts of the sample permittivity and also exhibit an extra gain equal to the square of the $Q$-boosting ratio to the changes in dielectric losses. Equivalent LC models for the ASRRs were developed and shown to closely match the behavior of the HFSS EM models. The optimum coupling conditions and pixel saturation were studied. The nonlinear effects of the $-g_m$ circuits inside the resonator and their impact on de-Qing the resonator, when driven by excessive power, were investigated, leading to an optimum input power. The effects of various noise sources, including ASRR device white and flicker noise, the supply noise, and the input phase noise, on the output phase noise were analyzed. Ultimately, these studies culminated in providing design guidelines for maximizing the SNR while minimizing the power consumption.


%

\appendices
\section{Output Phase Calculations}
\label{appa}
The output phase, $\varphi_{out}=\angle{S_{21}}$, and its derivatives are widely used in sensitivity and detection limits analysis, and are computed here. In (\ref{eq_s21}), $Z'$ is the impedance of the equivalent SRR, i.e., the parallel $R'L'C'$ resonator, given by 
\begin{equation}
    Z'(j\omega)=\frac{j\omega L'}{(1-\omega/\omega _0)(1+\omega/\omega _0)+j\omega L'/R'}
    \label{eq_rlc_imp}
\end{equation}
where $\omega _0$ is the resonance frequency. At the resonance frequency, $\Re(Z'(\omega_0))=R'$ and $\Im(Z'(\omega_0))=0$. For a small frequency offset from the resonance, i.e., $\omega _0+\Delta \omega$, the impedance can be approximated from
\begin{equation}
    Z'(j(\omega _0+\Delta \omega))\approx\frac{j(\omega _0 +\Delta \omega)L'}{-2\frac{\Delta \omega}{\omega _0}+j(\omega _0+\Delta \omega)\frac{L'}{R'}}
    \label{eq_rlc_imp_del}
\end{equation}

Hence, from (\ref{eq_rlc_imp_del}), the real and imaginary parts of the impedance can be approximated as follows
\begin{subequations}
\begin{align}
    &\Re{(Z'(\omega _0+\Delta \omega))}\approx R'\\
    &\Im{(Z'(\omega _0+\Delta \omega))}\approx -2\frac{\Delta \omega}{\omega_0}\frac{R'^2}{(\omega_0+\Delta \omega)L'}    
\end{align}
    \label{eq_rlc_re_im}
\end{subequations}

Equation (\ref{eq_rlc_re_im}b) can be used to find the derivative of the imaginary part as
\begin{equation}
    \frac{d\Im(Z')}{d\omega}=\lim_{\Delta \omega\to 0}\frac{\Im{(Z'(\omega _0+\Delta \omega))}}{\Delta\omega}\approx-2\frac{R'^2}{\omega_0^2L'}
    \label{eq_dim_dw}
\end{equation}

Since $\varphi_{out}=-\angle{(Z'+2Z_0)}$, one can derive expressions for the first quantity of interest, namely $\frac{d\varphi_{out}}{d\omega}$ as

\begin{equation}
    \frac{d\varphi_{out}}{d\omega}=-\frac{d}{d\omega}tan^{-1}\left(\frac{\Im(Z')}{\Re(Z')+2Z_0}\right)
    \label{eq_dphi_dw_1}
\end{equation}

Carrying out the differentiation and substituting (\ref{eq_rlc_re_im}a) and (\ref{eq_dim_dw}) into (\ref{eq_dphi_dw_1}) results in

\begin{align}
    \frac{d\varphi_{out}}{d\omega}&=\frac{-1}{1+(\frac{\Im(Z')}{R'+2Z_0})^2}\frac{1}{R'+2Z_0}\frac{d\Im(Z')}{d\omega}=\frac{2R'^2}{(R'+2Z_0)\omega_0^2L'}
    \label{eq_dphi_dw_2}
\end{align}

Using (\ref{eq_dphi_dw_2}), an expression for $\frac{d\varphi_{out}}{d\omega dR'}$ can be found as

\begin{equation}
    \frac{d\varphi_{out}}{d\omega dR'}=\frac{2R'(R'+4Z_0)}{L'\omega_0^2(R'+2Z_0)^2}=Q\frac{2}{\omega_0}\frac{R'+4Z_0}{(R'+2Z_0)^2}
    \label{eq_dphi_dwdr}
\end{equation}

Using (\ref{eq_srr}b), this quantity can be referred from the equivalent SRR to the actual SRR by first calculating

\begin{equation}
    \frac{R'}{R_{SRR}}=\frac{L'}{L_{SRR}}=\frac{M^2}{L_{SRR}^2}
\end{equation}
where $R_{SRR}$ is the equivalent parallel resistance of the resonator. Substituting $R'$ with $R_{SRR}$ in (\ref{eq_dphi_dwdr}) results in

\begin{equation}
    \frac{d\varphi_{out}}{d\omega dR_{SRR}}=Q\frac{2}{\omega_0}\frac{R'+4Z_0}{(R'+2Z_0)^2}\frac{M^2}{L_{SRR}^2}
    \label{eq_dphi_dwdr_2}    
\end{equation}




\section{Nonlinear Transconductance Calculations}
\label{appb}
The transconductance of a device in a $-g_m$ block can be divided into three cut-off, saturation, and triode regions and approximated as: 
\begin{align}
    &g_{m,off}=0\nonumber\\
    &g_{m,sat}\approx K\frac{W}{L}(V_{GS}-V_{TH})\nonumber\\
    &g_{m,tr}\approx K\frac{W}{L}(V_{DS})
\end{align}

Here, $V_{DS}=V_{DD}/2-(V_{ASRR}/2)sin\theta$ and $V_{GS}=V_{DD}/2+(V_{ASRR}/2)sin\theta$. If the voltage across the ASRR traverses all these three regions, as shown in Fig. \ref{fig:gm_comp}(\subref{fig:large_gm_sqr}), then $g_{m,avg}$ can be computed from

\begin{equation} \label{eq_gm_avg_int}
\begin{split}
    g_{m,avg}=\frac{1}{2\pi}\left[2\int_0^{\pi/2-\theta_C}(g_{m0}+K\frac{W}{L}\frac{V_{ASRR}}{2}sin\theta)d\theta+\right.\\
    \left.\int_{\pi/2-\theta_C}^{\pi/2+\theta_C}K\frac{W}{L}(\frac{V_{DD}}{2}-\frac{V_{ASRR}}{2}sin\theta)d\theta+\right.\\
    \left.2\int_\pi^{3\pi/2-\theta_C}(g_{m0}+K\frac{W}{L}\frac{V_{ASRR}}{2}sin\theta)d\theta\right]
\end{split}    
\end{equation}
where $\theta_C$ is the conduction angle in the triode region and is provided by

\begin{equation}
    \theta_C=
    \begin{cases}
        0&,\frac{V_{ASRR}}{2}\leq \frac{V_{TH}}{2}\\
        cos^{-1}\frac{V_{TH}}{V_{ASRR}}&,\frac{V_{ASRR}}{2}>\frac{V_{TH}}{2}
    \end{cases}
\end{equation}

A sine wave spends most of its time at its peaks, resulting in a sharp drop in $g_{m,avg}$ when the voltage swing enters the triode region. Carrying out the integrations in (\ref{eq_gm_avg_int}), $g_{m,avg}$ can be derived from the following expression
\begin{equation} \label{eq_gm_avg}
\begin{split}
    &g_{m,avg}=\\
    &\frac{1}{\pi}\left[g_{m0}(\pi-2\theta_C)+K\frac{W}{L}(\frac{V_{DD}}{2}\theta_C-\frac{V_{ASRR}}{2}sin\theta_C)\right]
\end{split}
\end{equation}

Equation (\ref{eq_gm_avg}) for $V_{ASRR}\gg V_{TH}$ can be very effectively approximated by replacing $\theta_C\approx\pi/2$ and thus a segmented expression for $g_{m,avg}$ can be established as
\begin{equation}
    g_{m,avg}\approx
    \begin{cases}
        g_{m0}&, V_{ASRR}\leq V_{TH}\\
        \frac{1}{\pi}K\frac{W}{L}(\frac{V_{DD}\pi}{4}-\frac{V_{ASRR}}{2})&,V_{ASRR}\gg V_{TH}
    \end{cases}
    \label{eq_gm_avg_aprx}
\end{equation}


\section*{Acknowledgment}

The authors would like to thank the Semiconductor Research Corporation (SRC) for funding this project.

\ifCLASSOPTIONcaptionsoff
  \newpage
\fi

\bibliographystyle{IEEEtran} 
\bibliography{ref} 

@article{fmartin_appphylet2003,
    author = {Mart\'in, F. and Bonache, J. and Falcone, F. and Sorolla, M. and Marqu\'es, R.},
    title = "{Split ring resonator-based left-handed coplanar waveguide}",
    journal = {Applied Physics Letters},
    volume = {83},
    number = {22},
    pages = {4652-4654},
    year = {2003},
    month = {12},
}

@ARTICLE{baena_srr_tmtt2005,
  author={Baena, J.D. and others},
  journal={IEEE Trans. on Microwave Theory and Techniques}, 
  title={Equivalent-circuit models for split-ring resonators and complementary split-ring resonators coupled to planar transmission lines}, 
  year={2005},
  volume={53},
  number={4},
  pages={1451-1461},
  doi={10.1109/TMTT.2005.845211}}

@ARTICLE{shang_srr_tmtt2013,
  author={Shang, Yang and Yu, Hao and Cai, Deyun and Ren, Junyan and Yeo, Kiat Seng},
  journal={IEEE Trans. on Microwave Theory and Techniques}, 
  title={Design of High-{Q} Millimeter-Wave Oscillator by Differential Transmission Line Loaded With Metamaterial Resonator in 65-nm {CMOS}}, 
  year={2013},
  volume={61},
  number={5},
  pages={1892-1902},
  doi={10.1109/TMTT.2013.2253489}}

@ARTICLE{rmarques_srr_tant2003,
  author={Marques, R. and Mesa, F. and Martel, J. and Medina, F.},
  journal={IEEE Trans. on Antennas and Propagation}, 
  title={Comparative analysis of edge- and broadside- coupled split ring resonators for metamaterial design - theory and experiments}, 
  year={2003},
  volume={51},
  number={10},
  pages={2572-2581},
  doi={10.1109/TAP.2003.817562}}

@ARTICLE{hillger_jssc2018,
  author={Hillger, Philipp and others},
  journal={IEEE J. of Solid-State Circuits}, 
  title={A 128-Pixel System-on-a-Chip for Real-Time Super-Resolution Terahertz Near-Field Imaging}, 
  year={2018},
  volume={53},
  number={12},
  pages={3599-3612},
  doi={10.1109/JSSC.2018.2878817}}

@INPROCEEDINGS{ameri_vlsi2019,
  author={Ameri, Ali and Zhang, Luya and Gharia, Asmaysinh and Niknejad, Ali M. and Anwar, Mekhail},
  booktitle={2019 Symposium on VLSI Circuits}, 
  title={A {114GHz} Biosensor with Integrated Dielectrophoresis for Single Cell Characterization}, 
  year={2019},
  volume={},
  number={},
  pages={C314-C315},
  doi={10.23919/VLSIC.2019.8778194}}

@ARTICLE{mitsunaka_jssc2016,
  author={Mitsunaka, Takeshi and others},
  journal={IEEE J. of Solid-State Circuits}, 
  title={{CMOS} Biosensor {IC} Focusing on Dielectric Relaxations of Biological Water With 120 and 60 {GHz} Oscillator Arrays}, 
  year={2016},
  volume={51},
  number={11},
  pages={2534-2544},
  doi={10.1109/JSSC.2016.2605001}}

@ARTICLE{tanaka_tcas2022,
  author={Tanaka, Akiyoshi and Chen, Guowei and Niitsu, Kiichi},
  journal={IEEE Trans. on Circ. and Syst. II: Express Briefs}, 
  title={A 4.5-m{W} 22-nm {CMOS} Label-Free Frequency-Shift 3 × 3 × 2 {3-D} Biosensor Array Using Vertically Stacked 60-{GHz} {LC} Oscillators}, 
  year={2022},
  volume={69},
  number={10},
  pages={4078-4082},
  doi={10.1109/TCSII.2022.3185542}}

@INPROCEEDINGS{ameri_isscc,
  author={Ameri, Ali and Chien, Jun-Chau and Niknejad, Ali},
  booktitle={2025 IEEE International Solid-State Circuits Conference (ISSCC)}, 
  title={20.10 A 200{GHz} 200-Pixel {2D} Near-Field Imager for Biomedical Applications}, 
  year={2025},
  volume={68},
  number={},
  pages={1-3},
  doi={10.1109/ISSCC49661.2025.10904598}}

@book{Darabi_rfic, place={Cambridge}, edition={2}, title={Radio Frequency Integrated Circuits and Systems}, publisher={Cambridge University Press}, author={Darabi, Hooman}, year={2020}}

@ARTICLE{wang_gasspec_tbiocas2018,
  author={Wang, Cheng and Perkins, Bradford and Wang, Zihan and Han, Ruonan},
  journal={IEEE Trans. on Biomedical Circ. and Syst.}, 
  title={Molecular Detection for Unconcentrated Gas With ppm Sensitivity Using 220-to-320-{GHz} Dual-Frequency-Comb Spectrometer in {CMOS}}, 
  year={2018},
  volume={12},
  number={3},
  pages={709-721},
  doi={10.1109/TBCAS.2018.2812818}}

@ARTICLE{wang_gasspec_jssc2021,
  author={Wang, Cheng and Yi, Xiang and Kim, Mina and Yang, Qingyu Ben and Han, Ruonan},
  journal={IEEE J. of Solid-State Circuits}, 
  title={A Terahertz Molecular Clock on {CMOS} Using High-Harmonic-Order Interrogation of Rotational Transition for Medium-/Long-Term Stability Enhancement}, 
  year={2021},
  volume={56},
  number={2},
  pages={566-580},
  doi={10.1109/JSSC.2020.3021335}}

@ARTICLE{laemmle_sens_tmtt2013,
  author={Laemmle, Benjamin and Schmalz, Klaus and Scheytt, J. Christoph and Weigel, Robert and Kissinger, Dietmar},
  journal={IEEE Trans. on Microwave Theory and Techniques}, 
  title={A 125-{GHz} Permittivity Sensor With Read-Out Circuit in a 250-nm {SiGe} {BiCMOS} Technology}, 
  year={2013},
  volume={61},
  number={5},
  pages={2185-2194},
  doi={10.1109/TMTT.2013.2253792}}

@ARTICLE{Grzyb_tmtt2017,
  author={Grzyb, Janusz and Heinemann, Bernd and Pfeiffer, Ullrich R.},
  journal={IEEE Trans. on Microwave Theory and Techniques}, 
  title={Solid-State Terahertz Superresolution Imaging Device in 130-nm {SiGe} {BiCMOS} Technology}, 
  year={2017},
  volume={65},
  number={11},
  pages={4357-4372},
  doi={10.1109/TMTT.2017.2684120}}

@Article{Ferrier_loc,
author ="Ferrier, Graham A. and others",
title  ="A microwave interferometric system for simultaneous actuation and detection of single biological cells",
journal  ="Lab Chip",
year  ="2009",
volume  ="9",
issue  ="23",
pages  ="3406-3412",
publisher  ="The Royal Society of Chemistry"}

@ARTICLE{chien_reactsens_jssc2016,
  author={Chien, Jun-Chau and Niknejad, Ali M.},
  journal={IEEE J. of Solid-State Circuits}, 
  title={Oscillator-Based Reactance Sensors With Injection Locking for High-Throughput Flow Cytometry Using Microwave Dielectric Spectroscopy}, 
  year={2016},
  volume={51},
  number={2},
  pages={457-472},
  doi={10.1109/JSSC.2015.2500362}}

@ARTICLE{Elhadidy_sens_tcas2015,
  author={Elhadidy, Osama and Shakib, Sherif and Krenek, Keith and Palermo, Samuel and Entesari, Kamran},
  journal={IEEE Trans. on Circ. and Syst. I: Regular Papers}, 
  title={A Wide-Band Fully-Integrated {CMOS} Ring-Oscillator {PLL}-Based Complex Dielectric Spectroscopy System}, 
  year={2015},
  volume={62},
  number={8},
  pages={1940-1949},
  doi={10.1109/TCSI.2015.2426960}}

@ARTICLE{cheng_imgr_mwtl2023,
  author={Cheng, Zong-Jun and others},
  journal={IEEE Microwave and Wireless Technology Letters}, 
  title={A 13-{GHz} “3-{D}” Near-Field Imager Employing Programmable Fringing Fields for Cancer Imaging}, 
  year={2023},
  volume={33},
  number={6},
  pages={931-934},
  doi={10.1109/LMWT.2023.3265584}}

@ARTICLE{hu_imgr_tbiocas2022,
  author={Hu, Kangping and Ho, Jason and Rosenstein, Jacob K.},
  journal={IEEE Trans. on Biomedical Circ. and Syst.}, 
  title={Super-Resolution Electrochemical Impedance Imaging With a 512 × 256 {CMOS} Sensor Array}, 
  year={2022},
  volume={16},
  number={4},
  pages={502-510},
  doi={10.1109/TBCAS.2022.3183856}}

@ARTICLE{Pfeiffer_uwavemag2019,
  author={Pfeiffer, Ullrich R. and others},
  journal={IEEE Microwave Magazine}, 
  title={Ex Vivo Breast Tumor Identification: Advances Toward a Silicon-Based Terahertz Near-Field Imaging Sensor}, 
  year={2019},
  volume={20},
  number={9},
  pages={32-46},
  doi={10.1109/MMM.2019.2922119}}

@article{huang_diffraction_cell2010,
  title={Breaking the diffraction barrier: super-resolution imaging of cells},
  author={Huang, Bo and Babcock, Hazen and Zhuang, Xiaowei},
  journal={Cell},
  volume={143},
  number={7},
  pages={1047--1058},
  year={2010},
  publisher={Elsevier}
}

@article{Abasi_cellsens_acsmeas2022,
  title={Bioelectrical impedance spectroscopy for monitoring mammalian cells and tissues under different frequency domains: A review},
  author={Abasi, Sara and others},
  journal={ACS measurement science au},
  volume={2},
  number={6},
  pages={495--516},
  year={2022},
  publisher={ACS Publications}
}

@ARTICLE{ameri_jssc25,
  author={Ameri, Ali and Chien, Jun-Chau and Niknejad, Ali M.},
  journal={IEEE Journal of Solid-State Circuits}, 
  title={Design and Implementation of a 200-{GHz} 200-Pixel 2-{D} Near-Field Imager}, 
  year={2025},
  volume={},
  number={},
  pages={1-16},
  doi={10.1109/JSSC.2025.3644751}}

@article{marques_physrevb2002,
  title={Role of bianisotropy in negative permeability and left-handed metamaterials},
  author={Marqu{\'e}s, Ricardo and Medina, Francisco and Rafii-El-Idrissi, Rachid},
  journal={Physical Review B},
  volume={65},
  number={14},
  pages={144440},
  year={2002},
  publisher={APS}
}

\end{document}